\documentclass[conference]{IEEEtran}
\IEEEoverridecommandlockouts
\usepackage{cite}
\usepackage{amsmath,amssymb,amsfonts}
\usepackage{algorithm}
\usepackage{algpseudocode}
\usepackage{acronym}
\usepackage{balance} %
\usepackage{bm}
\usepackage{booktabs}
\usepackage{graphicx}
\usepackage{import}
\usepackage{tikz}
\usepackage{wasysym} %
\usepackage{orcidlink}
\usepackage{pgfplots}
\usepackage{textcomp}
\usepackage{xcolor}
\usepackage{csvsimple}
\usepackage{gensymb}
\usepackage{comment}
\usepackage{subcaption}
\usepackage{soul}%
\usepackage{standalone}
\usepackage{svg}
\usepackage[normalem]{ulem}
\def\BibTeX{{\rm B\kern-.05em{\sc i\kern-.025em b}\kern-.08em
    T\kern-.1667em\lower.7ex\hbox{E}\kern-.125emX}}

\usepgfplotslibrary{external}
\usetikzlibrary{calc}
\usetikzlibrary{positioning}%
\usetikzlibrary{shapes}%
\usetikzlibrary{fit} %
\tikzset{>=latex} %

\newif\ifimporttikz
\newif\ifarxiv
\arxivtrue

\ifarxiv
  \importtikzfalse
\else
  \importtikztrue
\fi

\renewcommand{\thefootnote}{\arabic{footnote}}
\makeatletter
\renewcommand{\@makefnmark}{\textsuperscript{\thefootnote}}
\makeatother
\usepackage[symbol]{footmisc}

\acrodef{aegis}[AEGIS]{adaptive entropy-based Gaussian-mixture information synthesis}
\acrodef{agm}[AGM]{adaptive Gaussian mixture}
\acrodef{agmimm}[AGMIMM]{adaptive Gaussian mixture interacting multiple model}
\acrodef{car}[CAR]{constrained admissible region}
\acrodef{cb}[CB]{celestial barycenter}

\acrodef{cr3bp}[CR3BP]{circular restricted three-body problem}
\acrodef{cut}[CUT]{conjugate unscented transform}
\acrodef{cvm}[CvM]{Cram\'{e}r von Mises}
\acrodef{da}[DA]{differential algebra}
\acrodef{ecef}[ECEF]{Earth-centered Earth fixed}
\acrodef{eci}[ECI]{Earth-centered inertial}
\acrodef{ekf}[EKF]{extended Kalman filter}
\acrodef{elk}[ELK]{expected likelihood kernel}
\acrodef{emb}[EMB]{Earth-Moon barycentric}
\acrodef{fov}[FoV]{field-of-view}
\acrodefplural{fov}[FoVs]{fields-of-view}
\acrodef{geo}[GEO]{geostationary}
\acrodef{gm}[GM]{Gaussian mixture}
\acrodef{hotdogs}[HOTDOGS]{higher-order tensor-based deferral of Gaussian splitting}
\acrodef{imm}[IMM]{interacting multiple model}
\acrodef{ise}[ISE]{integral squared error}
\acrodef{ivp}[IVP]{initial value problem}
\acrodef{jms}[JMS]{jump Markov system}
\acrodef{lam}[LAM]{likelihood agreement measure}
\acrodef{kkt}[KKT]{Karush–Kuhn–Tucker}
\acrodef{mcr}[MCR]{maximal covariance ratio}
\acrodef{madem}[MaDEM]{Mahalanobis distance of the error in the mean}
\acrodef{nise}[NISE]{normalized integral squared error}
\acrodef{nrho}[NRHO]{near rectilinear halo orbit}
\acrodef{ode}[ODE]{ordinary differential equation}
\acrodef{pdf}[pdf]{probability density function}
\acrodef{sda}[SDA]{space domain awareness}
\acrodef{stm}[STM]{state transition matrix}
\acrodefplural{stm}[STMs]{state transition matrices}
\acrodef{stt}[STT]{state transition tensor}
\acrodef{srp}[SRP]{solar radiation pressure}
\acrodef{sut}[SUT]{scaled unscented transform}
\acrodef{ukf}[UKF]{unscented Kalman filter}
\acrodef{leo}[LEO]{low Earth orbit}

\acrodef{alodt}[ALoDT]{adaptive level of detail transform}
\acrodef{fos}[FOS]{first-order stretching}
\acrodef{sadl}[SADL]{statistical and deterministic linearization}
\acrodef{safos}[SA-FOS]{spherical-average first-order stretching}
\acrodef{sasos}[SA-SOS]{spherical-average second-order stretching}
\acrodef{sos}[SOS]{second-order stretching}
\acrodef{solc}[SOLC]{second-order linearization change}
\acrodef{us}[US]{uncertainty-scaled}
\acrodef{usfos}[US-FOS]{uncertainty-scaled first-order stretching}
\acrodef{ussolc}[US-SOLC]{uncertainty-scaled second-order linearization change}
\acrodef{wsasos}[W-SA-SOS]{whitened spherical-average second-order stretching}
\acrodef{wussadl}[W-US-SADL]{whitened uncertainty-scaled statistical and deterministic linearization}
\acrodef{wusfos}[W-US-FOS]{whitened uncertainty-scaled first-order stretching}
\acrodef{wussolc}[W-US-SOLC]{whitened uncertainty-scaled second-order linearization change}
\acrodef{wussos}[W-US-SOS]{whitened uncertainty-scaled second-order stretching}

\DeclareMathOperator*{\argmax}{arg\,max}
\newcommand{\gauss}[3]{\mathcal{N}\left(#1; \, #2, \, #3\right)}

\newif\ifshownotes
\shownotesfalse %
\newif\ifshowrevisionnotes
\showrevisionnotestrue %

\begin{document}

\title{Higher-Order Tensor-Based Deferral of Gaussian Splitting for Orbit Uncertainty Propagation
}

\author{\IEEEauthorblockN{G.\ Andrew Siciliano \orcidlink{0000-0001-9241-5251}, Keith A. LeGrand \orcidlink{0000-0003-3212-7201}}
\IEEEauthorblockA{\textit{School of Aeronautics and Astronautics} \\
\textit{Purdue University}\\
West Lafayette, IN, United States}
\and
\IEEEauthorblockN{Jackson Kulik \orcidlink{0009-0007-0740-0436}}
\IEEEauthorblockA{\textit{Department of Mechanical and Aerospace Engineering} \\
  \textit{Utah State University}\\
  Logan, UT, United States}
}

\maketitle

\begin{abstract}
Accurate propagation of orbital uncertainty is essential for a range of applications within space domain awareness.
Adaptive Gaussian mixture-based approaches offer tractable nonlinear uncertainty propagation through splitting mixands to increase resolution in areas of stronger nonlinearities, as well as by reducing mixands to prevent unnecessary computational effort.
Recent work introduced principled heuristics that incorporate information from the system dynamics and initial uncertainty to determine optimal directions for splitting.
This paper develops adaptive uncertainty propagation methods based on these robust splitting techniques.
A deferred splitting algorithm tightly integrated with higher-order splitting techniques is proposed and shown to offer substantial gains in computational efficiency without sacrificing accuracy.
Second-order propagation of mixand moments is also seen to improve accuracy while retaining significant computational savings from deferred splitting.
Different immediate and deferred splitting methods are compared in four representative test cases, including a low Earth orbit, a geostationary orbit, a Molniya orbit, and a multi-body cislunar orbit.
\end{abstract}


\section{Introduction}

\ifarxiv
\footnotetext{This work has been submitted to the IEEE for possible publication.
Copyright may be transferred without notice, after which this version may no longer be accessible.}
\fi
\begin{figure*}[htbp]
\centering
  \begin{tikzpicture}
    \node[anchor=north west] (img) at (0,0)
    {\includesvg[width=0.66\linewidth]{Figures/journal_sixth_draft.svg}};
  \end{tikzpicture}
  \caption{Deferred Gaussian mixture splitting in orbit uncertainty propagation.}
  \label{fig:splitting_graphic}
\end{figure*}
To ensure safe space operations, a core function of \ac{sda} is maintaining orbital state estimates of resident space objects.
Accurate descriptions of orbital uncertainty are crucial, especially with the ever-growing number of active and defunct space objects.
Though Gaussian representations offer convenient mathematical and computational properties, the nonlinearity of orbital dynamics can quickly render distributions non-Gaussian.
This issue is only compounded when observation opportunities are temporally sparse.

Existing work in nonlinear estimation has offered improved uncertainty propagation through better approximations of the transformed random variables' statistical moments.
These approaches include higher-order unscented transformation~\cite{adurthi2015ConjugateUnscentedTransformationBased,stojanovski2021HigherOrderUnscentedEstimator}, \ac{stt}~\cite{park2006NonlinearMappingGaussian,majji2008high,boone2024efficient}, and \ac{da}~\cite{servadio2020recursive,servadio2022MaximumPosterioriEstimation,acciarini2024NonlinearPropagationNonGaussian} methods, which provide higher-order approximations of the transformed mean, covariance, and higher-order moments.
The Gauss von Mises distribution introduced in \cite{horwood2014GaussMisesDistribution} exploits a cylindrical state space and permits evaluation of the general Bayesian filter for orbital applications.
Other work has focused on reducing the high computational burden of sampling-based methods using multi-fidelity approaches \cite{jones2019MultifidelityOrbitUncertainty}. Additional approaches rely on sparse-grid Polynomial Chaos or other Polynomial Chaos Expansion-type methods to approximate the probability density function over time \cite{jones2013NonlinearPropagationOrbit,jones2023incorporating,vittaldev2016SpacecraftUncertaintyPropagation}. Still others solve the Fokker-Planck equations for approximations of the probability density function through numerous distinct methods \cite{acciarini2024uncertainty, kumar2012nonlinear, adurthi2022estimation}.

\Acp{gm} have also proved a powerful tool for modeling nonlinearity-induced non-Gaussian uncertainty distributions while retaining many of the advantages of Gaussian distributions.
This paper focuses on \acp{gm} in large part because they offer a potentially more tractable representation of uncertainty compared to the outputs  of other nonlinear uncertainty propagation techniques.
The \acp{stt} can be used in coordination with these \acp{gm} to propagate mixands over time \cite{khatri2024HybridMethodUncertainty}.
The accuracy of a \ac{gm} nonlinear transformation approximation typically relies on the mixture resolution, where higher resolution in the output mixture leads to a better fit with the true non-Gaussian density.
In adaptive \ac{gm} methods, splitting Gaussians permits a local increase in the resolution of the distribution by adding more mixands.
To prevent computational waste, however, the directions of splitting must be chosen in a principled manner.
Various splitting criteria have been proposed based on linearization error~\cite{huber2008entropy}, differential entropy~\cite{demars2013EntropyBasedApproachUncertainty}, total mechanical energy variance~\cite{legrand2023BayesianAnglesOnlyCislunar, iannamorelli2025AdaptiveGaussianMixture}, field of view bounds~\cite{legrand2022SplitHappensImprecise}, linearization change~\cite{tuggle2018AutomatedSplittingGaussian}, and other measures of nonlinearity~\cite{losacco2024LowOrderAutomaticDomain, vittaldevMultidirectionalGaussianMixture}.
Recently, \cite{kulik2024NonlinearityUncertaintyInformed} introduced new families of higher-order splitting objectives for both identifying nonlinearity \cite{kulikLINEARCOVARIANCEFIDELITY} and optimizing the split direction for general nonlinear transformations.

This paper introduces new methods for orbit uncertainty propagation based on the higher-order splitting techniques proposed by the authors in \cite{kulik2024NonlinearityUncertaintyInformed}.
Many adaptive \ac{gm}-based techniques for uncertainty propagation perform \textit{immediate splitting}, wherein all splits of Gaussian mixands are made at the beginning of the considered time span.
This work instead considers \textit{deferred splitting} which, as depicted in Figure~\ref{fig:splitting_graphic}, delays splitting of a mixand until the impacts of the system's nonlinearities consistently exceed some specified tolerance.
A deferred splitting algorithm, named the \ac{hotdogs} algorithm, is developed that leverages transition matrix and tensor composition to avoid repeated costly \ac{ivp} solutions.
Second-order mean and covariance propagation is also shown to significantly improve \ac{gm} approximations of the uncertainty.
Furthermore, the specific structure of \ac{hotdogs} ensures the computational savings compared to immediate splitting are retained when incorporating second-order propagation of moments.
The performance of the uncertainty propagation using these methods is assessed using multiple figures of merit across four distinct, relevant cases spanning both two- and multi-body dynamical systems.
Preliminary results were presented in \cite{siciliano2025deferredSplitting}.
This work offers a more thorough treatment of the deferred splitting algorithm, its implementation, and its performance. The Gaussian splitting procedure itself is introduced in more detail, and the impact of the second-order propagation of moments is presented.

The remainder of this paper is organized as follows.
The mathematical problem formulation is developed in Section~\ref{sec:problem_formulation}.
The deferred splitting and uncertainty propagation methodology appears in Section~\ref{sec:methodology}.
Section~\ref{sec:background} provides background on Gaussian splitting, whitening transformations, and partial derivative tensors.
Numerical analysis results are discussed in Section~\ref{sec:results}, and conclusions are drawn in Section~\ref{sec:conclusion}.

\section{Problem Formulation}
\label{sec:problem_formulation}
This work considers the problem of approximating the \ac{pdf} $p_\mathbf{Z}(\mathbf{z})$ of some nonlinear function $\mathbf{z}=\mathbf{g}(\mathbf{x})$ of a random variable $\mathbf{x}$ with known distribution.
Throughout this work, lowercase bold symbols are used to denote vectors and vector functions, and uppercase bold symbols denote matrices and tensors.
Specific focus is given to the nonlinear transformation $\mathbf{g}$ representing the flow of a deterministic dynamical system $\bm{\varphi}_{\Delta t}$ over some time $\Delta t$ \cite{Katok_Hasselblatt_1995}.
The continuous-time dynamical system describing the evolution of a state vector $\mathbf{x}\in\mathbb{R}^n$ is specified by the system of ordinary differential equations
\begin{equation}
    \frac{\mathrm{d}\mathbf{x}}{\mathrm{d}t}=\mathbf{f}(\mathbf{x},  t)
\end{equation}
where $\mathbf{f}$ may depend explicitly on time.
Where possible to do so without confusion, the time argument is suppressed for brevity.
The flow map is then defined as the operator
\begin{align}
  \label{eq:flow-map}
  \bm{\varphi}_{\Delta t}(\mathbf{x}(t_{0}))=  \mathbf{x}(t_{0} + \Delta t) = \mathbf{x}(t_{0}) + \int_{t_{0}}^{t_{0} + \Delta t} \mathbf{f}(\mathbf{x}(\tau), \tau)d\tau
\end{align}
Though this work examines deterministic transformations of random variables, these heuristics and algorithms can be extended to filtering applications with process noise given suitable discrete-time approximations of the continuous-time process noise.

\section{Background}
\label{sec:background}
\subsection{Gaussian mixtures}
In Bayesian nonlinear estimation, \acp{gm} offer a flexible parameterization of non-Gaussian uncertainty and can approximate any density (with a finite number of discontinuities) to an arbitrary degree of accuracy.
A \ac{gm} consists of $L$ \textit{mixands}, where each mixand is a tuple $(w_{i}, \mathbf{m}_{i}, \mathbf{P}_{i})$ comprising a weight, mean, and covariance, respectively.
The \ac{gm} \ac{pdf} is then given by
\begin{align}
    p(\mathbf{x}) = \sum_{i=1}^{L} w_{i} \mathcal{N}(\mathbf{x}; \mathbf{m}_{i},\mathbf{P}_{i})
\end{align}
\ac{gm} state estimation methods leverage the property that Gaussianity of a random variable is preserved under linear transformation.
If the nonlinear map is quasi-linear within the local support of each mixand, then the linearly mapped mixand is approximately Gaussian, while allowing the composite mixture to be non-Gaussian \cite{sorenson1971RecursiveBayesianEstimation}.
If, on the other hand, the second- and higher-order behavior of the function is non-negligible over a mixand's effective support, the mixand can itself can be replaced with more, smaller variance mixands in a procedure known as \textit{Gaussian splitting}.
\subsection{Univariate splitting}
\begin{figure}[htbp]
\centering
  \tikzexternaldisable%
  \begin{tikzpicture}
  \tikzexternalenable%
    \node[anchor=north west] (img) at (0,0){\scalebox{1}{
  \ifimporttikz
    \tikzsetnextfilename{uni_split_ex}
    \import{Figures/univariate_split_example/}{uni_split_ex.tikz}
  \else
    \includegraphics{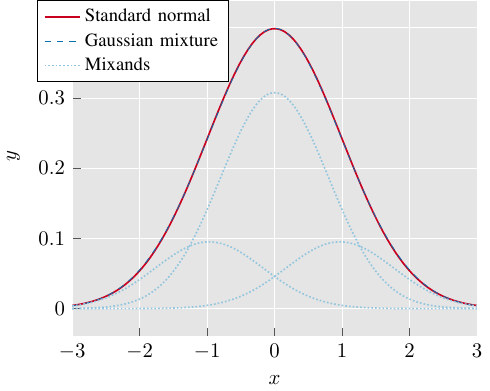}
  \fi
}};
  \end{tikzpicture}
  \caption{Three-component split of the standard univariate Gaussian.}
  \label{fig:uni_split_ex}
\end{figure}

The canonical splitting problem involves approximating the standard univariate Gaussian distribution $q(x)$ by a mixture of $L_{\mathrm{s}}$ mixands as
\begin{align}
  q(x) \approx
  \tilde{q}(x) = \sum_{i=1}^{L_{\mathrm{s}}} \tilde{w}_{j}
  \gauss{x}{\tilde{m}_{i}}{\tilde{\sigma}_{i}^{2}}
\end{align}
The best univariate split parameters are typically optimized to minimize some functional distance between the mixture approximation and the true density, subject to simplifying constraints, such as equally-spaced means and homoscedastic variances \cite{hanebeck2003ProgressiveBayesianEstimation, huber2008entropy}.
Introducing a regularization term $\lambda/L_{\mathrm{s}} \sum\tilde{\sigma}_{i}^{2}$ in the objective function penalizes large split mixand variances, which promotes increased mean separations and avoids the trivial solution of $L_{\mathrm{s}}$ identical mixands \cite{demars2013EntropyBasedApproachUncertainty}.
The splitting parameters are further constrained such that the mixture has the same overall mean and variance as the original Gaussian.
This paper adopts the optimization process described in \cite{kulik2024NonlinearityUncertaintyInformed}, which is omitted here for brevity.
Fortunately, this optimization process can be performed offline to build a library of solutions for varying $\lambda$ and $L_{\mathrm{s}}$, which then can be referenced efficiently online.
In Figure~\ref{fig:uni_split_ex}, a univariate splitting of the standard normal---zero mean and unity standard deviation---is demonstrated with $L_{\mathrm{s}} = 3$ post-split mixands and $\lambda = 0.0001$.

\subsection{Multivariate splitting}
\label{subsec:multivar_split}
A precomputed univariate splitting library is assumed to be available which contains entries corresponding to $L_{\mathrm{s}}$ mixands and that are given by $(\tilde{w}_i,\tilde{m}_i, \tilde{\sigma}_i)$.
From this, the multivariate splitting in the direction $\hat{\bm{\delta}}^*$ is performed as follows.
The means of each mixand are chosen to lie along the line parallel to $\hat{\bm{\delta}}^*$ with separation proportional to that of the original univariate splitting library as
\begin{equation}
  \mathbf{m}_i=\mathbf{m}+m_i\hat{\bm{\delta}}^*
  \label{eq:mean_placement}
\end{equation}
where $m_{i}$ is the corresponding library solution mean scaled by the directional reciprocal precision:
\begin{align}
  m_{i} = \frac{\tilde{m}_{i}}{(\hat{\bm{\delta}}^{*})^{\top} \mathbf{P}^{-1} \hat{\bm{\delta}}^{*}}
\end{align}
The covariance of each mixand is given by
\begin{equation}
    \mathbf{P}_i=\tilde{\sigma}_i^2\frac{\mathbf{P}-\sum_{i=1}^{L_{\mathrm{s}}} w_i(\mathbf{m}_i-\mathbf{m})(\mathbf{m}_i-\mathbf{m})^\top}{\sum_{i=1}^{L_{\mathrm{s}}} w_i\tilde{\sigma}_i^2}
\end{equation}
The weights of the multivariate splitting are identical to those from the original univariate splitting library.
Under this scheme, the mean and covariance of the overall Gaussian mixture is the same as the original distribution \cite{kulik2024NonlinearityUncertaintyInformed}.
This method for moment-matching multivariate splitting generalizes an existing method which assumed the original univariate splitting library had the same standard deviation for each mixand \cite{tuggle2018AutomatedSplittingGaussian}.
\subsection{Whitening transformations}
While whitening transformations are often employed in time series analysis to remove correlation over time, in this work whitening transformations are used to assist in defining uncertainty-weighted measurements of nonlinearity in a nonlinear function.
A whitening transformation associated with some zero-mean random variable $\mathbf{x}$ with covariance $\mathbf{P}$ is any linear transformation $\mathbf{W}$ such that the random variable
\begin{equation}
    \mathbf{y}=\mathbf{W}\mathbf{x}
\end{equation}
has identity covariance.
In particular, the inverse of any matrix square root of the covariance is a whitening transformation. One such square root is the Cholesky factor:
\begin{align}
    \mathbf{P}=\mathbf{L}\mathbf{L}^{\top} \implies \mathbf{W}= \mathbf{L}^{-1}
\end{align}

\subsection{Partial derivative tensors}
\label{subsec:pdts}
The Taylor series of the nonlinear function $\mathbf{g}$ truncated to second-order are utilized throughout this work.
Two primary approaches are taken.
In the first, the first-order term in the series is used in the traditional manner for covariance propagation in an extended Kalman filter.
The second-order term is employed to approximate the first-order term evaluated at different reference points and to analyze the error in linear covariance propagation to aid in Gaussian mixture splitting.
In the other approach, the second-order term is also used in the propagation of the moments.
The second-order Taylor series of $\mathbf{g}$ is written as
\begin{align}
  \mathbf{g}(\mathbf{m}_{x} + \bm{\delta})
  &=
  \mathbf{g}( \mathbf{m}_{x})
  +
  \mathbf{G}\bm{\delta}
  +
  \frac{1}{2}
  \mathbf{G}^{(2)} \bm{\delta}^{2}
  + \mathcal{O}(\bm{\delta}^3)
\end{align}
where the Jacobian and second-order partial derivative tensors are defined as
\begin{align}
  \label{eq:first_and_second_pdts}
  \mathbf{G}=\left.\frac{\partial\mathbf{g}}{\partial \mathbf{x}}\right|_{\mathbf{x}=\mathbf{m}_x}  \qquad \textrm{and} \qquad
  \mathbf{G}^{(2)} & =\left.\frac{\partial^2 \mathbf{g}}{\partial \mathbf{x}^2}\right|_{\mathbf{x}=\mathbf{m}_x}
\end{align}
The shorthand $\mathbf{G}^{(2)} \bm{\delta}^2$ is used to denote the double contraction of the tensor $\mathbf{G}^{(2)}$ with two copies of $\bm{\delta}$, defined more explicitly such that the $i$th component of the output is given by
\begin{align}
(\mathbf{G}^{(2)}\bm{\delta}^2)^i=G_{j, k}^i \delta^j \delta^k & =\left.\frac{\partial g^i}{\partial x^j \partial x^k}\right|_{\mathbf{x}=\mathbf{m}_x} \delta^j \delta^k
\end{align}
where summation over the repeated indices is implied according to Einstein summation convention.

Note also that the linear approximation of the Jacobian evaluated at another point near the mean is given by
\begin{equation}
\label{eq:stm_linear_approximation}
    \left.\frac{\partial g^i}{\partial x^j}\right|_{\mathbf{x}=\mathbf{m}_x+\bm{\delta}}=\left.\frac{\partial g^i}{\partial x^j}\right|_{\mathbf{x}=\mathbf{m}_x}+\left.\frac{\partial^2 g^i}{\partial x^j\partial x^k}\right|_{\mathbf{x}=\mathbf{m}_x} \delta^k + \mathcal{O}(\bm{\delta}^2)
\end{equation}
By this approximation, a state transition matrix (which is the Jacobian of the solution flow) can be obtained for points near the mean without repeated \ac{ivp} solutions, as shown in the following section.

\section{Methodology} \label{sec:methodology}
Accurate \ac{gm} uncertainty propagation relies on good local approximations of the solution flow around each mixand.
The validity of these approximations typically depends on the scale of the initial uncertainty relative to the level of nonlinearity of the function through which the uncertainty propagation is to be performed.
Previous work from the authors developed splitting criteria that measure the relative level of uncertainty and nonlinearity associated with a given mixand and work to identify heuristic directions for mixand splitting \cite{kulik2024NonlinearityUncertaintyInformed}.
These splitting directions are chosen to most effectively reduce the relative level of uncertainty and nonlinearity associated with the resulting mixands after the split.
Selection of the splitting direction\textemdash represented by the unit vector $\hat{\bm{\delta}}^{*}$\textemdash can be formulated in terms of an optimization subject to an induced norm constraint:
\begin{align}
  \label{eq:unit_constrained_opt}
  \bm{\delta}^*
  &=
  \argmax_{\Vert\bm{\delta}\Vert_\mathbf{A}=1}\mathcal{F}(\mathbf{g},\bm{\delta}; \Theta)\\
  \label{eq:unit_constraint}
  \hat{\bm{\delta}}^* &=
  \frac{  \bm{\delta}^*}{ \| \bm{\delta}^*\|_{2}}
\end{align}
where $\mathcal{F}$ is some functional acting on functions $\mathbf{g}$ and vectors $\bm{\delta}$ with parameters $\Theta$ to produce a non-negative real number, and $\Vert\bm{\delta}\Vert_\mathbf{A}=\sqrt{\bm{\delta}^{\top}\mathbf{A}\bm{\delta}}$ denotes the norm induced by the symmetric positive definite matrix $\mathbf{A}$.
Table~\ref{tab:summary_splitting_heuristics} summarizes the metrics introduced in \cite{kulik2024NonlinearityUncertaintyInformed} and employed in this work.

The basic variants rely on one of four following measures of nonlinearity in the objective functional.
\textit{\Ac{fos}}-based objectives do not directly measure nonlinearity, but rather the local linear stretching of space around the mixand mean.
\textit{\Ac{sos}}-based objectives measure nonlinearity via the second-order term in the Taylor expansion locally around the given mixand mean.
\textit{\Ac{solc}}-based objectives measure nonlinearity through how much the Jacobian of the nonlinear function changes as the expansion is carried out around another nearby reference point.
This objective also examines the second-order coefficient in the Taylor series expansion about the mixand mean, but uses a single contraction with a deviation vector rather than a double contraction as is the case with \ac{sos}-based methods.
\textit{\Ac{sadl}}-based methods employ a deterministic and statistical linearization about the mixand mean, comparing the two as a means of quantifying nonlinearity. The deterministic linearization is simply the Jacobian of the nonlinear function about the mixand mean.

Given an input mean $\mathbf{m}_{x}$ and covariance $\mathbf{P}_{x}$, the statistical linearization matrix $\mathbf{G}^{(\textrm{SL})}$ is obtained from a weighted least squares fit of an affine model
\begin{align}
    \mathbf{z} \approx \mathbf{G}^{(\textrm{SL})}\mathbf{x} + \mathbf{b}
\end{align}
with solution
\begin{align}
  \label{eq:statistical_linearization_matrix}
\mathbf{G}^{(\textrm{SL})}=\left(\mathbf{P}_{x z}\right)^{\top}\left(\mathbf{P}_x\right)^{-1} \text { and } \mathbf{b}=\mathbf{m}_z-\mathbf{G}^{(\textrm{SL})} \mathbf{m}_x
\end{align}
where the output mean $\mathbf{m}_{z}$, cross-covariance $\mathbf{P}_{xz}$, and covariance $\mathbf{P}_{z}$ may be obtained from some sigma point transformation \cite{huber2011AdaptiveGaussianMixture}.

These four variants describe how nonlinearity is accounted for in the objective functional of \eqref{eq:unit_constrained_opt}.
In addition to nonlinearity, the metrics and splitting direction selections should also depend on the scale of uncertainty present in a given mixand.
This consideration is incorporated into the constraint of the optimization from \eqref{eq:unit_constrained_opt}.
Metrics and splitting direction heuristics that optimize the nonlinearity functional over a surface of equal initial mixand likelihood are referred to as \textit{uncertainty-scaled}, as they consider the level of nonlinearity given by equally likely inputs instead of inputs equidistant from the mixand mean.
This means that the mixand's squared-Mahalanobis distance metric is employed in the optimization constraint and the norm employed is induced by the precision (inverse of covariance) matrix associated with the mixand.

Two other final modifications to the aforementioned metrics and algorithms are now considered.
The first is to average the effects of nonlinearity over the sphere (or equal probability ellipsoid) to identify directions that on average align well with strong nonlinearity.
The second is to perform a whitening transformation on the outputs (and also an inverse whitening transformation to inputs in the case of \ac{wussolc}) of the functions appearing inside a norm in the objective functional.
This serves to change the objective functional from a 2-norm to a Mahalanobis distance associated with the final mixand covariance.
This also has the added benefit of rendering the metric and splitting direction invariant under linear changes of coordinates, including changes to units.
The whitening transformation can be obtained from the final covariance of the initial mixand as determined by linear covariance propagation or the unscented transform.

The output whitening transformation strategy introduced in~\cite{kulik2024NonlinearityUncertaintyInformed} involved recomputing the linear approximation-based output covariance at each split recursion, such that the whitening matrix adjusted with the decreasing input mixand covariances.
While this approach provides the desired properties of measure invariance under linear coordinate changes and interpretability as a Mahalanobis distance, the resulting heuristic does not necessarily monotonically decrease with each split, which is undesirable.
This behavior is illustrated in the example in Appendix~\ref{sec:whitening_ex}.
Instead, this work proposes the use of the output covariance of the original Gaussian for the whitening of itself and all child mixands.
Thus, if the index pair $i,j$ refers to the $j$\textsuperscript{th} resulting child mixand from splitting parent mixand $i$, then the whitening transformation used for that child and all its descendants is simply the parent mixand's whitening transformation:
\begin{align}
    \label{eq:whitening_transformation}
    \mathbf{W}_{i,j}(t) =
    \mathbf{W}_{i}(t) =
    \mathbf{P}_{i,z}^{-1/2}(t)
\end{align}
where $\mathbf{P}_{i,z}^{-1/2}(t)$ is the reference non-split Gaussian's linearly-mapped output covariance at time $t$.
Performing whitening with the parent mixand covariance ensures that the \ac{wussolc} metric always decreases after a split while maintaining coordinate invariance of the metric.
Employing the post-split mixand covariance for metric calculation may lead to counterintuitive increases in \ac{wussolc} after splits despite increases in accuracy of the propagated distribution and is avoided in \ac{hotdogs}.

In the context of orbit uncertainty propagation, the nonlinear function under consideration is the flow map~\eqref{eq:flow-map} associated with a dynamical system: in particular, perturbed two-body and multi-body dynamics for near-Earth and cislunar environments, respectively.
The Jacobian and second-order partial derivative tensor of the flow map are known as the \ac{stm} and \ac{stt}, respectively.
The ordinary differential equations associated with the \ac{stm} and \ac{stt} can be obtained by taking partial derivatives of \eqref{eq:flow-map} and exchanging the order of derivatives under the assumption of sufficient regularity of the flow-map (continuity of second-order derivatives in time and phase space) which comes from a sufficiently smooth vector field.
These equations, known as the first- and second-order variational equations, are given in Einstein notation as
\begin{align}
    \label{eqn:variational}
    \frac{d\Phi^i_j(t_f,t_0)}{dt}&=\frac{\partial {\mathrm{f}}^{i}(\mathbf{x})}{\partial x^{l}} \Phi^l_j(t_f,t_0)\\
    \label{eqn:variational_2}
    \frac{d\Psi^i_{j,k}(t_f,t_0)}{dt}
    &=\frac{\partial^2 {\mathrm{f}}^{i}(\mathbf{x})}{\partial x^{l} \partial x^{q}} \Phi^l_j(t_f,t_0)\Phi^q_k(t_f,t_0)\\
    &+\frac{\partial {\mathrm{f}}^{i}(\mathbf{x})}{\partial x^{l}} \Psi^l_{j,k}(t_f,t_0)\nonumber\\
    \Phi^i_j(t_0,t_0)&=\delta^i_{j}\\
    \Psi^i_{j,k}(t_0,t_0)&=0
\end{align}
where $\delta^i_j$ denotes the Kronecker delta, and $\boldsymbol{\Phi}, \boldsymbol{\Psi}$ denote the \ac{stm} and second-order \ac{stt} respectively.
Alternatively, finite differencing \cite{pellegrini2016computation} or differential algebra techniques \cite{valli2013nonlinear} may be employed to obtain the partial derivative tensors associated with the flow-map at a particular reference state and time-of-flight.
Thus, in the context of dynamical uncertainty propagation, if $\mathbf{P}_{i}(t_{0})$ is the non-split reference input covariance, the whitening transformation~\eqref{eq:whitening_transformation} used for all child mixands is
\begin{align}
    \mathbf{W}_{i}(t) = \left(\bm{\Phi}(t,t_{0})\mathbf{P}_{i}(t_{0})\bm{\Phi}^{\top}(t,t_{0})\right)^{-1/2}
\end{align}

\begin{table}[htbp]
    \centering
    \caption{Summary of splitting heuristics, where the quantities listed in the constraint column are constrained to equal unity.}
    \label{tab:summary_splitting_heuristics}
      \setlength{\tabcolsep}{2pt}
    \begin{tabular}{ccc}
    \toprule[2pt]%
         Heuristic & Objective, $\mathcal{F}(\mathbf{g},\mathbf{\bm{\delta}}; \Theta)$  & Constraint \\
      \midrule
         maxvar & $\Vert \mathbf{\bm{\delta}} \Vert$ & $\Vert \mathbf{\bm{\delta}} \Vert_{\mathbf{P}_{x}^{-1}}$ \\
      \acs{fos}   & $\left\Vert\mathbf{G}\mathbf{\bm{\delta}}\right\Vert_2$ & $\|\mathbf{\bm{\delta}}\|_2$\\
      \acs{sos}   & $\Vert\mathbf{G}^{(2)}\mathbf{\bm{\delta}}^2\Vert_2$ & $\|\mathbf{\bm{\delta}}\|_2$\\
      \acs{solc} \cite{tuggle2018AutomatedSplittingGaussian,tuggle2020ModelSelectionGaussian} & $\Vert\mathbf{G}^{(2)}\mathbf{\bm{\delta}}\Vert_F$ & $\|\mathbf{\bm{\delta}}\|_2$\\
      \acs{sadl}  & $\Vert(\mathbf{G}^{(\textrm{SL})}-\mathbf{G})\mathbf{\bm{\delta}}\Vert_2$ & $\|\mathbf{\bm{\delta}}\|_2$\\
      \acs{usfos} & $\left\Vert\mathbf{G}\mathbf{\bm{\delta}}\right\Vert_2$ & $\Vert\mathbf{\bm{\delta}}\Vert_{\mathbf{P}^{-1}_x}$\\
      \acs{ussolc} \cite{tuggle2018AutomatedSplittingGaussian,tuggle2020ModelSelectionGaussian}& $\Vert\mathbf{G}^{(2)}\mathbf{\bm{\delta}}\Vert_F$ & $\Vert\mathbf{\bm{\delta}}\Vert_{\mathbf{P}^{-1}_x}$\\
      \acs{safos} & $\int\limits_{\mathbf{u}^{\top}\mathbf{P}_x^{-1} \mathbf{u}}\!\!\!\!\!\! (\mathbf{u}^{\top}\mathbf{\bm{\delta}})^{2}\Vert \mathbf{G}\mathbf{u}\Vert_{2}^{2} \mathrm{d} \varphi(\mathbf{u})$ & $\|\mathbf{\bm{\delta}}\|_2$
\\
      \acs{sasos} & $\int\limits_{\mathbf{u}^{\top}\mathbf{P}_x^{-1} \mathbf{u}}\!\!\!\!\!\! (\mathbf{u}^{\top}\mathbf{\bm{\delta}})^{2}\Vert \mathbf{G}^{(2)}\mathbf{u}^2\Vert_{2}^{2} \mathrm{d} \varphi(\mathbf{u})$ & $\|\mathbf{\bm{\delta}}\|_2$\\
      \acs{wussos} & $\Vert \mathbf{G}^{(2)}\mathbf{\bm{\delta}}^2\Vert_{\mathbf{W} \mathbf{W}^\top}$ & $\Vert\mathbf{\bm{\delta}}\Vert_{\mathbf{P}^{-1}_x}$\\
      \acs{wussolc} & $\Vert {\mathbf{W}}(\mathbf{G}^{(2)}\mathbf{\bm{\delta}})\mathbf{P}_x^{1/2}\Vert_{F}^2$ &$\Vert\mathbf{\bm{\delta}}\Vert_{\mathbf{P}^{-1}_x}$\\
      \acs{wussadl} & $\Vert \mathbf{W}(\mathbf{G}^{(\textrm{SL})}-\mathbf{G})\mathbf{\bm{\delta}}\Vert_{2}$ & $\Vert\mathbf{\bm{\delta}}\Vert_{\mathbf{P}^{-1}_x}$\\
      \acs{wsasos} & $\int\limits_{\mathbf{u}^{\top}\mathbf{P}_x^{-1} \mathbf{u}}\!\!\!\!\!\! (\mathbf{u}^{\top}\mathbf{\bm{\delta}})^{2}\Vert \mathbf{G}^{(2)}\mathbf{u}^2\Vert_{\mathbf{W} \mathbf{W}^{\top}}^{2} \mathrm{d} \varphi(\mathbf{u})$ & $\|\mathbf{\bm{\delta}}\|_2$\\
    \bottomrule[2pt]
    \end{tabular}
\end{table}

\subsection{Deferred Splitting}
For nonlinear \acp{ode}, numerical solution of the solution flow, \ac{stm}, and \ac{stt} can be computationally expensive, especially for long integration time spans.
These costs scale with the number of mixands, and thus it is desirable to reduce the quantity and time spans of these numerical solutions.
A \textit{deferred split} is defined as an intermediate splitting of a Gaussian that does not occur at the initial time; rather, the splits are performed as needed when a given threshold is exceeded.
The idea was introduced in \cite{demars2013EntropyBasedApproachUncertainty} in conjunction with an entropy-based split criterion and maximum-variance--based split direction.
Inspired by this concept, this section introduces a deferred splitting algorithm named \ac{hotdogs} that is tightly integrated with the higher-order splitting techniques described earlier.
Further, three variations of the \ac{hotdogs} algorithm are developed that significantly reduce the computational requirements of adaptive uncertainty propagation.

Consider the mapping of
\begin{align}
  p(\mathbf{x}(t_{0})) = \sum_{i=1}^{L} w_{i} \gauss{\mathbf{x}}{\mathbf{m}_{i}(t_{0})}{\mathbf{P}_{i}(t_{0})}
\end{align}
to $p(\mathbf{x}(t_{f}))$.
Using linear covariance approximation, the propagated mixands are given by
\begin{align}
  \mathbf{m}_{i}(t_{f}) &= \bm{\varphi}_{t_{f}-t_{0}} (\mathbf{m}_{i}(t_{0}))\\
  \label{eq:covariance_prop}
  \mathbf{P}_{i}(t_{f}) &= \bm{\Phi}(t_{f},t_{0}) \mathbf{P}_{i}(t_{0}) \bm{\Phi}^{\top}(t_{f},t_{0})
\end{align}
The underlying premise of deferred splitting is to maintain a fixed-size distribution up to the intermediate time $t_{s}\leq t_{f}$ where the nonlinearity of any mixand~$i$ meets some tolerance~$\epsilon$.
The splitting time selection is given by the optimization problem
\begin{align}
  \label{eq:optimal_split_time}
  t_{s}=&\max t, \quad \textrm{s.t.} \quad  t\leq t_{f} \quad \textrm{and} \\
        & \!\!\!\!\!\!\!\! w\mathcal{F}(\bm{\varphi}_{t-t_{0}}, \bm{\delta}; \mathbf{x}(t_{0}), \mathbf{m}(t), \mathbf{P}(t), \!\bm{\Phi}(t,t_{0}), \!\bm{\Psi}(t,t_{0}),\! \mathbf{W}(t))\!<\!\epsilon \nonumber
\end{align}
where the requisite parameters of $\mathcal{F}$ are written explicitly to highlight this dependence and the mixand index is omitted for brevity.
Then, by splitting the intermediate solution mixand $(w,\mathbf{m}(t_{s}), \mathbf{P}(t_{s}))$, the children mixands need only be mapped through the ``shorter'' flow $\varphi_{t_{f}-t_{s}}$.
Considering both the mixand weight $w$ and the objective $\mathcal{F}$ in this optimization emphasizes the effect on the full distribution; strong nonlinearities acting on a mixand with very low weight may be less impactful overall than weaker nonlinearities on a higher weight component.
The recursive application of this process comprises \ac{hotdogs}, outlined in Algorithm~\ref{alg:intermediate_splitting}.
Within this algorithm, three variants of increasing fidelity---DS-1, DS-2, and DS-3---are proposed, and are discussed later in further detail.
The size of the output mixture can be controlled by specifying a maximum recursion depth and/or a minimum mixand weight value, $w_{\min}$.

In practice, the optimal split time in~\eqref{eq:optimal_split_time} is solved approximately and trivially by restricting the candidate times to a discrete set $\{t_{0}, t_{1}, \hdots, t_{f}\}$.
The intermediate means, \acp{stm}, and \acp{stt} associated with these times are readily obtained as opportunistic outputs of the numerical integration solution.
Given the intermediate \acp{stm} $\bm{\Phi}(t,t_{0})$, the covariance matrix $\mathbf{P}(t)$ is also readily obtained via~\eqref{eq:covariance_prop}.
It should also be noted that the objective functional is not monotonically increasing in time, so the final time where the tolerance is satisfied is chosen as the splitting time.
Otherwise, splitting at the first tolerance violation could lead to wasted effort in propagating additional mixands if the nonlinearities later drop back below the desired tolerance.
Furthermore, because the time derivatives given in~\eqref{eqn:variational} and~\eqref{eqn:variational_2} also depend on the evolution of the state, the means, \acp{stm}, and \acp{stt} must be jointly integrated.

\begin{algorithm}
  \caption{\texttt{\ac{hotdogs}}}
\label{alg:intermediate_splitting}
    \begin{algorithmic}
        \Require $w,\mathbf{m}(t), \mathbf{P}(t), \bm{\Phi}(t,t_{0}),\bm{\Psi}(t,t_{0}), \mathbf{W}(t), t_{f}, t_{0}$
        \If {$w<w_{\min}$}
        \State \Return $\{(w,\mathbf{m}(t_{f}),\mathbf{P}(t_{f}))\}$
        \EndIf
        \State $t_{s}=\max t, \quad \textrm{s.t.} \quad  t\leq t_{f} \,\, \textrm{and}$ \\
        $ \, \, w\mathcal{F}(\bm{\varphi}_{t\!-\!t_{0}}, \!\bm{\delta}^{*}; \mathbf{x}(t_{0}), \mathbf{m}(t), \mathbf{P}(t), \!\bm{\Phi}(t,t_{0}), \!\bm{\Psi}(t,t_{0}), \!\mathbf{W}\!(t))\!<\!\epsilon$
        \If {$t_{s}=t_{f}$}
        \State \Return $\{(w,\mathbf{m}(t_{f}),\mathbf{P}(t_{f}))\}$
        \EndIf
        \State Obtain $\bm{\Psi}(t,t_{s})$ via~\eqref{eqn:inverse_cocycle2} and $\bm{\Phi}(t,t_{s})$ via~\eqref{eq:intermediate_stm}
        \State $\{w_{i},\mathbf{m}_{i}(t_{s}), \mathbf{P}_{i}(t_{s}) \}_{i=1}^{L_{\mathrm{s}}}\gets \texttt{split}(w,\mathbf{m}(t_{s}), \mathbf{P}(t_{s}))$
        \For {$i=1,\hdots,L_{\mathrm{s}}$}
        \State Obtain $\bm{\Psi}_{i}(t,t_{s})$ and $\bm{\Phi}_{i}(t,t_{s})$ by DS-1, DS-2, or DS-3
        \State $\mathbf{P}_{i}(t) = \bm{\Phi}_{i}(t,t_{s}) \mathbf{P}_{i}(t_{s})\left(\bm{\Phi}_{i}(t,t_{s})\right)^{\top}$
        \State $\mathbf{m}_{i}(t) = \bm{\varphi}_{t-t_{s}}( \mathbf{m}_{i}(t_{s}))$
        \EndFor
        \State \Return $\left\lbrace \bigcup \texttt{\ac{hotdogs}}(\right.$ \\
        $\qquad \qquad \left. \!\!\!\!w_{i},\mathbf{m}_{i}(t), \mathbf{P}_{i}(t), \bm{\Phi}_{i}(t,t_{s}),\bm{\Psi}_{i}(t,t_{s}), \mathbf{W}(t), t_{f}, t_{s})\right\rbrace$
    \end{algorithmic}
\end{algorithm}

If $t_{s}=t_{f}$, $w < w_{\min}$, or the maximum recursion depth has been reached, then no further splitting is performed for that mixand.
Otherwise, the mixand is split along the direction maximizing the objective functional $\mathcal{F}$.
For this optimization, the parameters needed are not the original $\bm{\Phi}(t,t_{0})$ and $\bm{\Psi}(t,t_{0})$, but rather the split time-referenced $\bm{\Phi}(t,t_{s})$ and $\bm{\Psi}(t,t_{s})$.
Fortunately, these tensors can be obtained without numerical integration by leveraging the composition properties of flow maps and their associated \acp{stm} and \acp{stt}.

The \ac{stm} starting at $t_{s}$ is readily obtained as
\begin{align}
    \label{eq:intermediate_stm}
    \bm{\Phi}(t_{f},t_{s}) = \bm{\Phi}(t_{f},t_{0}) \bm{\Phi}^{-1}(t_{s},t_{0})
\end{align}
Using tensor notation as introduced in Section~\ref{subsec:pdts}, \Ac{stt} composition is given by
\begin{align}
\label{eq:stt_composition}
    &\Psi^{i}_{j,k} (t_{f},t_{0}) = \\
    &\qquad \Psi^{i}_{q,l}(t_{f},t_{s}) \Phi^{q}_{j}(t_{s},t_{0})\Phi_{k}^{l}(t_{s},t_{0}) + \Phi^{i}_{l}(t_{f},t_{s}) \Psi^{l}_{j,k}(t_{s},t_{0}) \nonumber
\end{align}
where $\bm{\Psi}(t_{f},t_{s})$ is the quantity of interest.
Solving~\eqref{eq:stt_composition}, the \ac{stt} starting at $t_{s}$ is given by
\begin{align}
\label{eqn:inverse_cocycle2}
    \Psi^i_{j,k}(t_{f},t_{s})
    =[\Psi^i_{l,m}(t_{f},t_{0})-\Phi^i_{q}(t_{f},t_{s})\Psi^q_{l,m}(t_{s},t_{0})] \nonumber \\
    \cdot (\Phi^{-1})^l_j(t_{s},t_{0})(\Phi^{-1})^m_k(t_{s},t_{0})
\end{align}
To compute this in a numerically stable and efficient fashion, let
\begin{equation}
    A^i_{l,m}=\Psi^i_{l,m}(t,0)-\Phi^i_{q}(t,\tau)\Psi^q_{l,m}(\tau,0)
\end{equation}
and
\begin{equation}
    B^i_{j,m}=A^i_{l,m}(\Phi^{-1})^l_j(t_{s},t_{0})
\end{equation}
For fixed indices $i,m$, the vector $B^i_{:,m}$ is obtained by solving the linear system
\begin{equation}
\Phi^{\top}(t_{s},t_{0})B^i_{:,m}=A^i_{:,m}
\end{equation}
using LU decomposition and forward/backward substitution.
Every element of the tensor $\mathbf{B}$ is obtained in this manner.
The tensor $\boldsymbol{\Psi}(t_{f},t_{s})$ is then obtained in similar fashion using the solutions of the set of linear systems
\begin{equation}
\Phi^{\top}(t_{s},t_{0})\Psi^i_{j,:}=B^i_{j,:}
\end{equation}

This aforementioned inverse composition process provides the necessary tensors for optimizing the split direction without repeating costly numerical integration.
A challenge arises in that these tensors represent an expansion only about the trajectory $\mathbf{m}(t)$ and are generally not equal to the unknown expansions about the split mixand mean trajectories $\mathbf{m}_{i}(t)$.
The remainder of this section details three methods for obtaining the split mixand expansion terms
$\bm{\Phi}_{i}(t,t_{s}) = \frac{\partial \bm{\varphi}_{t-t_{s}}}{\partial \mathbf{x}^2}\big|_{\mathbf{x}=\mathbf{m}_{i}(t_{s})}$
and
$\bm{\Psi}_{i}(t,t_{s}) = \frac{\partial^{2} \bm{\varphi}_{t-t_{s}}}{\partial \mathbf{x}}\big|_{\mathbf{x}=\mathbf{m}_{i}(t_{s})}$ within \ac{hotdogs}.
These methods are enumerated and abbreviated as DS-1, DS-2, and DS-3.

The DS-1 method approximates the \acp{stt} as being constant over input deviations; that is,  $\bm{\Psi}_{i}(t,t_{s}) = \bm{\Psi}(t,t_{s})$.
This approximation avoids the most expensive operation of numerically integrating the new \acp{stt} and is generally accurate when third- and higher-order variations of the flow are small over the mixand's effective support.
The DS-1 method also avoids numerical integration of the new \acp{stm} and instead approximates them by employing the linear approximation~\eqref{eq:stm_linear_approximation} with respect to the reference mixand mean.
\begin{align}
    \big[{\Phi}_{i'}(t,t_{s})\big]^{l}_{j}
    &\approx
    \big[{\Phi}_{i}(t,t_{s})\big]^{l}_{j} \nonumber \\
    &+
    \big[{\Psi}_{i}(t,t_{s})\big]^{l}_{j,k} \left( \big[{m}_{i'}(t_{s})\big]^{k} - \big[{m}_{i}(t_{s})\big]^{k} \right)
\end{align}
By these two approximations, the DS-1 method is very fast but incurs the greatest accuracy penalty of the proposed methods.
Caution should be exercised when applying the DS-1 method, particularly when the mixand separations are large.
In these cases, the linear \ac{stm} approximation can deteriorate, and may not preserve the symplectic property of the \ac{stm} in Hamiltonian dynamical systems.

\begin{figure*}[htbp]
    \centering
    \scalebox{0.88}{
  \ifimporttikz
    \tikzsetnextfilename{hotdogs_block_diagram}
    \import{Figures/}{hotdogs_block_diagram.tikz}
  \else
    \includegraphics{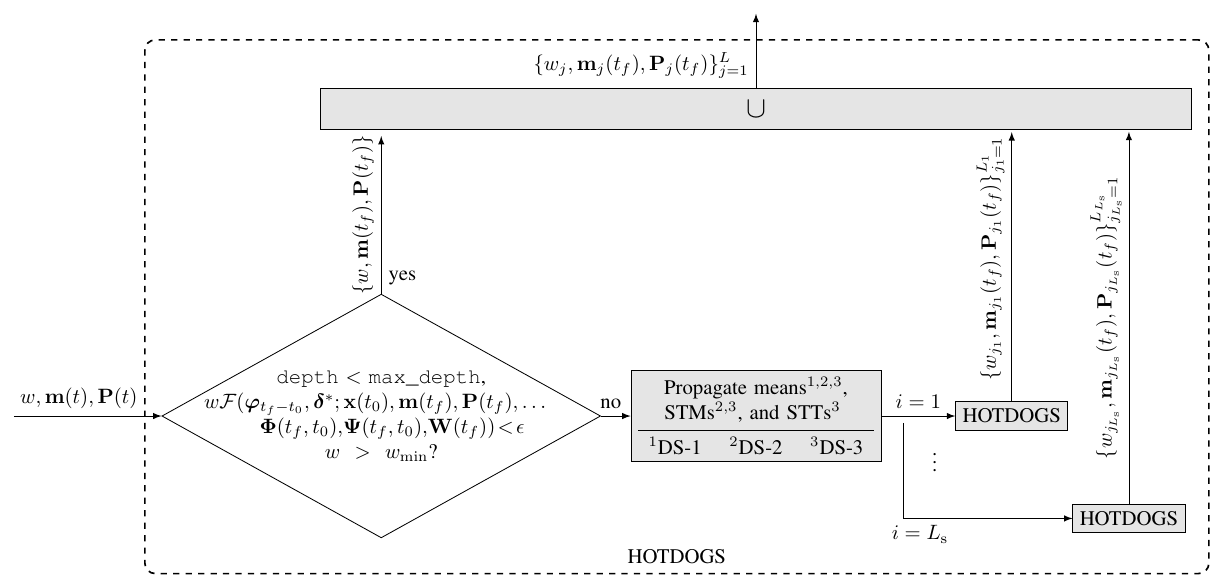}
  \fi
}
    \caption{\ac{hotdogs} block diagram.}
    \label{fig:hotdogs_flow_chart}
\end{figure*}

The DS-2 method similarly applies a constant \ac{stt} approximation but recomputes the new \acp{stm} using numerical integration.
By this approach, DS-2 offers improved accuracy by avoiding the most severe approximation employed by DS-1.
The DS-3 method offers the highest accuracy at increased computational cost by recomputing the \acp{stm} and the \acp{stt} for each split mixand.
The full algorithm is depicted in Figure~\ref{fig:hotdogs_flow_chart}.

Several additional techniques can exploit the advantageous emergent structure to further reduce computational effort.
When splitting into an odd number of mixands, the central resultant mixand retains the same means over time as the original mixand under first-order propagation.
So, recomputation of the means, \acp{stm}, and \acp{stt} can be avoided for that central component.
Furthermore, with first-order propagation of the means and covariances, the \acp{stt} are used only for splitting heuristics.
Therefore, once the split recursion depth limit is reached, the \ac{stt} computation can be skipped entirely for that final depth.

\subsection{Second-Order Moment Propagation}
Linear covariance analysis has seen frequent usage due to its simplicity and computational ease; however, the first-order mean and covariance propagations which lend it the epithet inherently lead to challenges in highly nonlinear scenarios.
The \ac{stt}-based propagation in \cite{park2006NonlinearMappingGaussian} retains higher-order terms from the Taylor series, which can better handle such stressing systems.
As second-order \acp{stt} are incorporated in all of the second-order splitting methods---including the deferred splitting \ac{hotdogs} algorithm---they are readily exploited to extend mixand moment propagation to second-order.
Truncating the Taylor series at the second order and applying to the propagation of a given mixand's moments extends~\eqref{eq:covariance_prop} to
\begin{align}\label{eq:second_order_prop}
  \big[\delta m_{i}(t_f)\big]^{s} &= \frac{1}{2} \big[{\Psi}_{i}(t_f, t_0)\big]^s_{q,r} \big[{P}_{i}(t_0)\big]^{q,r}\\
  \big[{m}_{i}(t_{f})\big]^{u} &= \big[{\varphi}_{t_{f}-t_{0}} ({m}_{i}(t_{0}))\big]^{u} + \big[\delta {m}_{i}(t_f)\big]^{u} \\
  C_i(t_0) &= \big[{P}_{i}(t_0)\big]^{n,o} \big[{P}_{i}(t_0)\big]^{p,q} + \big[{P}_{i}(t_0)\big]^{n,p}  \nonumber \\
  &\cdot \big[{P}_{i}(t_0)\big]^{o,q} + \big[{P}_{i}(t_0)\big]^{n,q} \big[{P}_{i}(t_0)\big]^{o,p}\\
  \big[{P}_{i}(t_{f}) \big]^{j,k} &= \big[ {\Phi}(t_{f},t_{0}) \big]^{j}_{l} \big[ {\Phi}(t_{f},t_{0}) \big]^{k}_{m} \big[{P}_i(t_{0})\big]^{l,m}  \nonumber\\
  &- \big[\delta {m}_{i}(t_f)\big]^{j} \big[\delta {m}_{i}(t_f)\big]^{k}  \nonumber\\
  &+ \frac{1}{4} \big[{\Psi}_{i}(t_f, t_0)\big]^{j}_{n,o} \big[{\Psi}_{i}(t_f, t_0)\big]^{k}_{p,q} C_i(t_0)
\end{align}
This higher-order moment approximation can significantly improve overall uncertainty propagation accuracy compared to standard first-order approximations, especially in the presence of strong nonlinearity.
Furthermore, for the splitting methods with precomputed \acp{stt}, this significant improvement in performance can be achieved with little added computational effort.

\subsection{Practical Considerations}
Though the considered splitting criteria and split direction selection are presented as general constrained optimization problems, in practice all but the \ac{sos}-based heuristics can be solved as eigenvalue or singular value problems.
The \ac{sos}-based methods, on the other hand, employ shifted power iteration-based strategies.
While conservative shift parameter selection methods exist that guarantee convergence, they can result in a slow rate of convergence.
Appendix~\ref{sec:shift_calculation} presents a new choice of shift parameter and derives an upper bound which still guarantees convergence while significantly increasing the rate of convergence compared to existing choices.
An open-source Python implementation of this new shift parameter computation, as well as all of the splitting heuristics in this paper, is available in the PyEst library at \url{https://github.com/scope-lab/pyest}.
Computation of the \acp{stm} and \acp{stt} is performed using differential algebra and a custom, optimized version of DACEyPy \url{https://github.com/scope-lab/daceypy} that has been modified from the original version \cite{doi:10.2514/6.2018-0398} to enable efficient extraction of intermediate-time maps.

\section{Results}
\label{sec:results}
The proposed uncertainty propagation methods are analyzed in four scenarios: a geostationary orbit, a low Earth orbit, a Molniya orbit, and an Earth-Moon southern $\mathrm{L}_{1}$ halo orbit.
The general form of the equations of motion for dynamical models considered here is
\begin{equation}
\mathbf{f}(\mathbf{x}, t) =
\begin{bmatrix}
\mathbf{v}(t) \\[1mm]
\mathbf{a}(\mathbf{r}, \mathbf{v}, t)
\end{bmatrix}, \quad
\mathbf{x} =
\begin{bmatrix} \mathbf{r} \\ \mathbf{v} \end{bmatrix} \in \mathbb{R}^6, \
\mathbf{r,v,a} \in \mathbb{R}^3
\end{equation}
where  $\mathbf{r} $ is the position,  $\mathbf{v}$ the velocity, and $\mathbf{a}$ is the  acceleration.
The perturbed two-body equations of motion are expressed in the Earth-centered inertial (ECI) J2000 frame, with acceleration
\begin{equation}
\begin{aligned}
\mathbf{a}(\mathbf{r},\mathbf{v}, t) =\;&
\mathbf{a}_{2\mathrm{B}}(\mathbf{r})
+ \varepsilon_{J_2}\,\mathbf{a}_{J_2}(\mathbf{r})
+ \varepsilon_{\mathrm{drag}}\,\mathbf{a}_{\mathrm{drag}}(\mathbf{r},\mathbf{v})
\\
&\quad + \varepsilon_{\mathrm{SRP}}\,\mathbf{a}_{\mathrm{SRP}}(\mathbf{r}, t)
 + \varepsilon_{\leftmoon}\,\mathbf{a}_{\mathrm{\leftmoon}}(\mathbf{r}, t),
\end{aligned}
\end{equation}
where $\mathbf{a}_{2\mathrm{B}}$ is the central-body gravitational acceleration. The perturbing accelerations
$\mathbf{a}_{J_2}$, $\mathbf{a}_{\mathrm{drag}}$, $\mathbf{a}_{\mathrm{SRP}}$, and $\mathbf{a}_{\mathrm{\leftmoon}}$
correspond to, respectively, the Earth's oblateness, atmospheric drag (with exponential density and a non-rotating atmosphere), the \ac{srp} under the cannonball assumption, and the Moon gravitational perturbation; each contribution is enabled through the binary variables
$\varepsilon_{J_2},\varepsilon_{\mathrm{drag}},\varepsilon_{\mathrm{SRP}}, \varepsilon_{\leftmoon}\in\{0,1\}$ depending on its relevance to the specific orbital regime in question\cite{Curtis2014Chapter12}.

In the cislunar application, the considered dynamical model accounts for the gravitational action of multiple bodies in the \ac{emb} J2000 inertial frame, with acceleration
\begin{equation}\label{eq:nbody}
\mathbf{a}(\mathbf{r}, t) =
\sum_{i\in\mathcal{B}}
\mathbf{a}_{\mathrm{B},i}(\mathbf{r})
+ \mathbf{a}_{\mathrm{EMB/CB}}(t)
+ \mathbf{a}_{\mathrm{SRP}}(\mathbf{r}, t),
\end{equation}
where $\mathcal{B}$ is the selected set of bodies (Earth, Moon, Sun, and Jupiter, retrieved from ephemerides) and $\mathbf{a}_{\mathrm{B},i}$ the associated gravitational contribution.
The term $\mathbf{a}_{\mathrm{EMB/CB}}(t)$ accounts for the \ac{emb} acceleration with respect to the \ac{cb}, while $\mathbf{a}_{\mathrm{SRP}}$ is the solar radiation pressure acceleration, again modeled via the cannonball approximation. Collectively, these contributions define the total acceleration $\mathbf{a}$ that governs the flow map  operator via $\mathbf{f}$.

In all test cases considered in this section, the initial distributions are Gaussian.
From each parent mixand, an individual split produces $L_{\mathrm{s}}=3$ child mixands using the procedure detailed in Section~\ref{subsec:multivar_split}.
In the immediate splitting case, the split recursion is carried out to the chosen maximum recursion depth of three, such that the final number of mixands equals~27.

The same maximum number and depth of splits is applied to the deferred splitting methods, with the key difference being that these splits may generally occur at different times during the propagation period.

Splitting to the maximum recursion depth eliminates sensitivity to differences in the methods' split tolerances, permitting a more balanced comparison.
The splitting directions are selected according to the heuristics first presented in \cite{kulik2024NonlinearityUncertaintyInformed} and discussed in Section~\ref{sec:methodology}, as well as two other heuristics considered for comparison.
Namely, the ``maxvar'' method splits along the direction of highest variance, while the \ac{alodt} splits along the principal direction in which the deviation between propagated sigma points and their linear regression fit is maximized \cite{faubel2010FurtherImprovementAdaptive}.

The orbital state is defined in terms of Cartesian position and velocity coordinates.
The true distribution is represented by propagating forward 10,000 Monte Carlo samples through the considered time spans of 7 days for the near Earth orbits and 14 days for the cislunar orbit.
The \ac{geo} orbit is propagated with initial covariance
\begin{align}
    \mathbf{P}(t_{0}) = \text{diag}([10, 10, 1, 3, 3, 0.01])^2
\end{align}
where the position is in kilometers and the velocity in meters per second.
The low Earth orbit selected has a semi-major axis of $6,725$ kilometers, an eccentricity of $0.0004$, an inclination of $97.3\degree$, and an argument of periapsis of $144.8\degree$.
The Molniya orbit considered is an example of a highly elliptical orbit, which represents a more challenging uncertainty propagation problem compared to the \ac{leo} or \ac{geo} cases.
The Molniya orbit is characterized by its high eccentricity of $0.737$, semi-major axis of 26,553 kilometers, inclination of $63.4\degree$, and argument of periapsis of $270\degree$.
The \ac{leo} and Molniya scenarios share a common initial covariance of
  \begin{align}
    \mathbf{P}(t_{0}) = \text{diag}([0.01,\, 0.01, \, 0.01, \, 0.1, \, 0.1,  \, 0.1])^2
\end{align}
where the listed standard deviation values are in kilometers and meters per second.
These three orbits consider perturbed two-body dynamics: the \ac{geo} case considers \ac{srp} and the Moon's gravity, the \ac{leo} case incorporates drag and $J_2$ effects, and the Molniya case includes drag, $J_2$ effects, and \ac{srp}.
For the purposes of demonstrating the capture of uncertainty growth over time, a simple exponential atmospheric model is considered sufficient for the drag computation.

\begin{figure*}
\centering
\begin{subcaptiongroup}
  \subcaptionlistentry{throwaway}
  \label{fig:geo_traj}
  \centering
  \tikzexternaldisable%
  \begin{tikzpicture}
  \tikzexternalenable%
  \node[anchor=north west] (img) at (0,0){\scalebox{0.6}{
  \ifimporttikz
    \tikzsetnextfilename{geo_traj}
    \import{Figures/geo_trajectory/}{geo_traj.tikz}
  \else
    \includegraphics{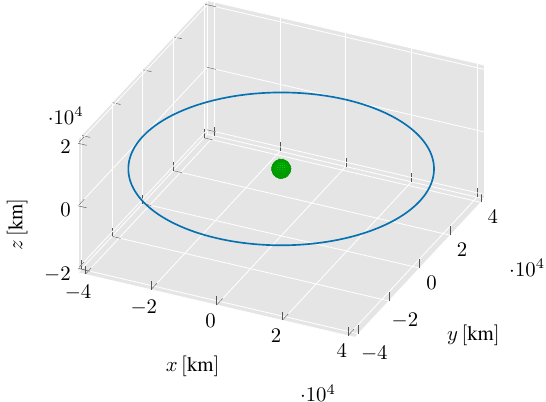}
  \fi
}};
    \node at (0.4,-0.6) {\captiontext*{}};
  \end{tikzpicture}
  \subcaptionlistentry{throwaway}
  \label{fig:leo_traj}
  \centering
  \tikzexternaldisable%
  \begin{tikzpicture}
  \tikzexternalenable%
  \node[anchor=north west] (img) at (0,0){\scalebox{0.6}{
  \ifimporttikz
    \tikzsetnextfilename{leo_traj}
    \import{Figures/leo_trajectory/}{leo_traj.tikz}
  \else
    \includegraphics{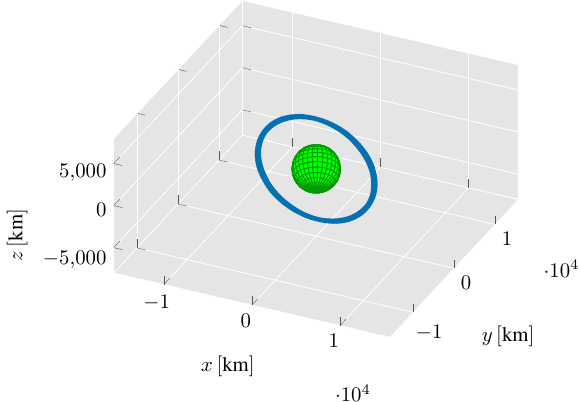}
  \fi
}};
    \node at (0.4,-0.6) {\captiontext*{}};
  \end{tikzpicture}
  \subcaptionlistentry{throwaway}
  \label{fig:molniya_traj}
  \centering
  \tikzexternaldisable%
  \begin{tikzpicture}
  \tikzexternalenable%
  \node[anchor=north west] (img) at (0,0){\scalebox{0.6}{
  \ifimporttikz
    \tikzsetnextfilename{molniya_traj}
    \import{Figures/Molniya_trajectory/}{molniya_traj.tikz}
  \else
    \includegraphics{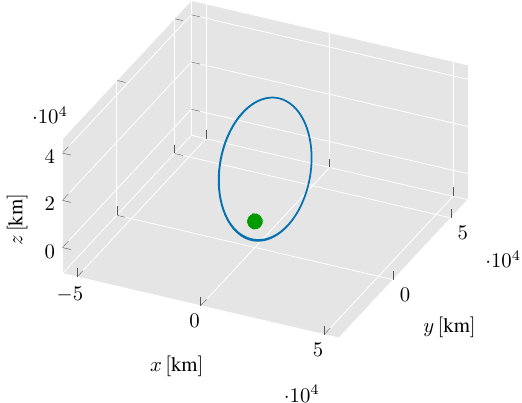}
  \fi
}};
    \node at (0.4,-0.6) {\captiontext*{}};
  \end{tikzpicture}
  \subcaptionlistentry{throwaway}
  \label{fig:cislunar_traj}
  \centering
  \tikzexternaldisable%
  \begin{tikzpicture}
  \tikzexternalenable%
  \node[anchor=north west] (img) at (0,0){\scalebox{0.6}{
  \ifimporttikz
    \tikzsetnextfilename{cislunar_traj}
    \import{Figures/cislunar_trajectory/}{cislunar_traj.tikz}
  \else
    \includegraphics{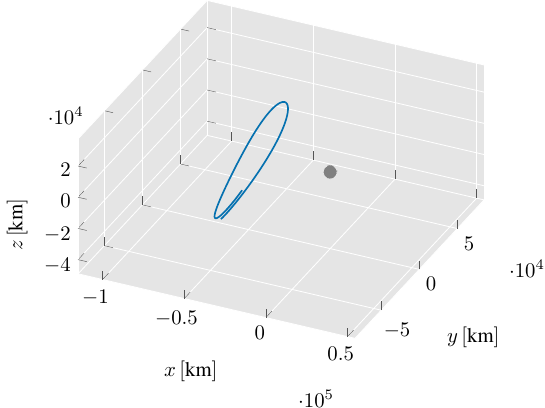}
  \fi
}};
    \node at (0.4,-0.6) {\captiontext*{}};
  \end{tikzpicture}
\end{subcaptiongroup}
\captionsetup{subrefformat=parens}
\caption{\subref{fig:geo_traj} Geostationary orbit, \subref{fig:leo_traj} low Earth orbit, \subref{fig:molniya_traj} Molniya orbit, and \subref{fig:cislunar_traj} multi-body southern $\mathrm{L}_{1}$ halo orbit.}
\label{fig:3_cases}
\end{figure*}

The fourth scenario involves the multi-body Earth–Moon system, where the motion of the satellite is influenced by the gravitational attraction of the Earth, Moon, Sun, and Jupiter as well as \ac{srp}.
The high-fidelity multi-body dynamics appropriate for the cislunar regime, as defined in~\eqref{eq:nbody}, are employed, and the reference orbit is chosen to be a southern $\mathrm{L}_{1}$ halo orbit, with initial conditions in kilometers and meters per second in the \ac{emb} J2000 frame:
\begin{align}
    \mathbf{r}(t_{0}) &= \begin{bmatrix}
        -62809.1, -8862.7, -40873.9
    \end{bmatrix}\\
    \mathbf{v}(t_{0}) &= \begin{bmatrix}
        -22.791, 202.241, -41.647
    \end{bmatrix}
\end{align}
The corresponding epoch for this state is 12:00:00 UTC on November 17, 2025, while the initial covariance is
\begin{align}
    \mathbf{P}(t_{0}) = \text{diag}([40, 4, 40, 0.01, 0.01, 0.01])^2
\end{align}
and the distribution is propagated for 14 days.
The four orbits are shown in Figure~\ref{fig:3_cases}, where the green central body represents the Earth and the gray represents the Moon.

\subsection{Splitting Objective Comparison} \label{sec:results_baseline}
The impact of splitting objective choice on the overall uncertainty propagation accuracy is first examined using immediate splitting.
Accuracy is assessed by three figures of merit---the \ac{madem}, \ac{mcr}, and \ac{cvm} norm---designed to compare different statistical features between the Gaussian mixture representation and the Monte Carlo samples, where the Monte Carlo result is treated as the truth distribution \cite{kulik2024NonlinearityUncertaintyInformed}.
The \ac{madem}, \ac{mcr}, and \ac{cvm} norm measures can be interpreted as the agreement of the approximation's mean, covariance, and overall distribution shape, respectively, with the sample population.
Identical distributions yield values of zero \ac{madem}, unity \ac{mcr}, and zero \ac{cvm} norm, so results decreasing towards these values indicate approximations which more closely match the corresponding attributes of the Monte Carlo samples.

The \ac{madem} metric examines the difference between the means of the \acp{pdf} $p$ and $p'$---the \ac{gm} and Monte Carlo distributions respectively.
\begin{align}
    \left \Vert \boldsymbol{\mu}_{x} - \boldsymbol{\mu}_{x}' \right \Vert_{\mathbf{P}_{x}^{-1}}
\end{align}
A lower \ac{madem} indicates greater agreement between the estimated means of the approximate and empirical densities.

The \ac{mcr} extends this idea to the second moments by measuring the disagreement of the 1-$\sigma$ covariance ellipsoids.
The maximum and minimum values of the ratio
\begin{align} \label{eq:mcr_ratio}
    \frac{\mathbf{x}^{\top} \mathbf{P}_{x}^{-1} \mathbf{x}}{\mathbf{x}^{\top} (\mathbf{P}_{x}')^{-1} \mathbf{x}}
\end{align}
quantify this by comparing the difference between the directions of greatest and least agreement between the covariance ellipsoids.
The generalized eigenvalue problem
\begin{align}
    \mathbf{P}^{-1}_{x} \mathbf{x} = \lambda (\mathbf{P}_{x}')^{-1} \mathbf{x}
\end{align}
returns the eigenvalues which give the stationary values of the ratio \eqref{eq:mcr_ratio}.
The \ac{mcr} is then $\text{max}(1 / \text{min}_i~\lambda_i, \text{max}_i~\lambda_i)$: a measure of the deviation in covariances between the two distributions.
An \ac{mcr} of unity implies perfect agreement between the distributions' covariances, while larger values indicate increasing disagreement.

More holistically, the \ac{cvm} norm is used to measure the goodness-of-fit between the \ac{gm} approximation and the $N$ Monte Carlo samples.
Considering the $j$th marginal of the $m$-dimensional random variable $\mathbf{x}$, the univariate \ac{cvm} metric between the cumulative marginal distribution $F^{*}(x^{j})$ and an $N$-sample empirical distribution $F_{N}(x^{j})$ is
\begin{align}
    \omega_{j}^{2} = \int_{-\infty}^{\infty}\left(F^{*}(x^{j}) - F_{N}(x^{j})\right)^{2}\text{d}F^{*}(x^{j})
\end{align}
To quantify the overall goodness-of-fit in the multivariate case, the 2-norm is taken of the $m$ marginal \ac{cvm} metrics arrayed in a vector as
\begin{align}
    ||\boldsymbol{\omega}^{2}||_{2} = \left \Vert \begin{bmatrix}
        \omega_{1}^{2} \,\dots \,\omega_{m}^{2}
    \end{bmatrix}\right \Vert_{2}
\end{align}
and referred to as the \ac{cvm} norm.

The results of uncertainty propagation for the \ac{geo} orbit are listed in Table~\ref{tab:GEO_baseline_results}.
Notably, \ac{fos} here performs comparably to its more computationally straining counterparts in \ac{sos}, \ac{solc}, and \ac{sadl}.
Similarly, \ac{usfos} remains competitive with \ac{ussolc} and the heuristics using whitening transformations.
This supports the observation in \cite{kulik2024NonlinearityUncertaintyInformed} that first-order stretching approaches are well suited for many astrodynamics applications, as the direction of greatest linear stretching often aligns closely with that of strongest nonlinearity in two- and  multi-body dynamics.
Overall, most heuristics perform well and in similar manners to each other; this indicates the direction which induces the strongest non-Gaussianities manifests such that it is apparent to each of the disparate methods used to select splitting direction.
Furthermore, as the uncertainty scaling does not noticeably improve performance in this case, this direction is likely closely aligned with a direction of high initial uncertainty.
The similarity between \ac{ussolc} and \ac{wussolc}---and in general between the heuristics and their whitened variants---is unsurprising; the chosen units and coordinates do not place undue stress upon the system.
Hence, there is little room for the whitened methods to differentiate themselves.
Conversely, the spherical-averaging--based heuristics struggle in this scenario, indicating that primary directions of strongest nonlinearity are not closely aligned.
\begin{table}[htbp]
    \caption{Geostationary orbit case.}
    \label{tab:GEO_baseline_results}
    \centering
    \begin{tabular}{lrrrr}
      \toprule[2pt]%
      Method & MaDEM & CvM norm & MCR\\
      \midrule
      \begin{filecontents*}{data/twobody_geo_results.csv}
Method,ELK,MHEM,CvMnorm,MCR,Mixands
maxvar,4403573.833,996.1898,23.9547,1432.4240,27
FOS,38414785.22,0.1160,0.9497,1.0976,27
SOS,38649411.67,0.1175,0.9496,1.0976,27
SOLC,38646033.79,0.1175,0.9496,1.0976,27
SADL,38392843.21,0.1160,0.9497,1.0976,27
US-FOS,38418081.59,0.1160,0.9497,1.0976,27
US-SOLC,38661070.37,0.1175,0.9496,1.0976,27
SA-FOS,3597961.479,325.2943,23.9546,467.7446,27
SA-SOS,4044054.482,191.4911,23.9544,275.3496,27
W-US-SOS,38384784.89,0.1160,0.9496,1.0976,27
W-US-SOLC,38384726.43,0.1160,0.9496,1.0976,27
ALoDT,38405085.98,0.1160,0.9497,1.0976,27
W-SA-SOS,3975659.319,190.7221,23.9544,274.2442,27
W-US-SADL-S,38400952.73,0.1160,0.9496,1.0976,27
W-US-SADL-D,38400952.73,0.1160,0.9496,1.0976,27
\end{filecontents*}
\csvreader[head to column names]{data/twobody_geo_results.csv}{}
      {\Method & \MHEM & \CvMnorm & \MCR\\}\\[-1em]%
      \bottomrule
    \end{tabular}
\end{table}

The trends for the \ac{leo} case in Table~\ref{tab:LEO_baseline_results} are quite similar to the \ac{geo} case.
With the exception of maxvar and the spherical averaging-based methods---and in this case the slightly degraded performance with \ac{alodt} as well---all other heuristics perform very well.
Indeed, the relatively poor performance of maxvar in general is to be expected due to its lack of consideration for the dynamics of the system.
Here, the uncertainty scaling seems to provide a slight improvement, but changes on this scale can often be due more to stochasticity from comparing against a sampled truth distribution than to any physical factors.
\begin{table}[htbp]
    \caption{Low Earth orbit case.}
    \label{tab:LEO_baseline_results}
    \centering
    \begin{tabular}{lrrrr}
      \toprule[2pt]%
      Method & MaDEM & CvM norm & MCR\\
      \midrule
      \begin{filecontents*}{data/twobody_leo_results.csv}
Method,ELK,MHEM,CvMnorm,MCR,Mixands
maxvar,3.55E+11,6.2285,2.8791,8.8980,27
FOS,6.54E+11,0.1070,0.6953,1.0605,27
SOS,6.55E+11,0.1070,0.6953,1.0605,27
SOLC,6.55E+11,0.1070,0.6953,1.0605,27
SADL,7.61E+11,0.1036,0.6869,1.0591,27
US-FOS,6.11E+11,0.0986,0.6784,1.0600,27
US-SOLC,6.12E+11,0.0986,0.6784,1.0600,27
SA-FOS,4.19E+11,2.1121,2.7272,3.1624,27
SA-SOS,4.96E+11,1.3108,2.5247,2.1176,27
W-US-SOS,6.11E+11,0.0986,0.6784,1.0600,27
W-US-SOLC,6.11E+11,0.0986,0.6784,1.0600,27
ALoDT,4.43E+11,0.2800,1.2379,1.1014,27
W-SA-SOS,3.69E+11,1.3102,2.5244,2.1170,27
W-US-SADL-S,6.90E+11,0.1027,0.6846,1.0596,27
W-US-SADL-D,6.90E+11,0.1027,0.6846,1.0596,27
\end{filecontents*}
\csvreader[head to column names]{data/twobody_leo_results.csv}{}
      {\Method & \MHEM & \CvMnorm & \MCR\\}\\[-1em]%
      \bottomrule
    \end{tabular}
\end{table}

The Molniya orbit proves to be overall more stressing than the \ac{geo} and \ac{leo} cases, as demonstrated in Table~\ref{tab:molniya_baseline_results}.
While the same splitting heuristics which exhibited strong results before continue to stand out in terms of \ac{madem} and \ac{mcr}, the accompanying \ac{cvm} norm is very high in this case, which indicates a poor goodness-of-fit between the Gaussian mixtures and the truth distribution.
However, the \ac{madem} and \ac{mcr} reveal that the first moment and the direction of maximal covariance are captured well by the transformed \ac{gm} approximation.
This suggests that the linear covariance propagation fails to capture the growth of the covariance in other directions.
As this poor fit is a consequence of the propagation as opposed to the splitting itself, the high \ac{cvm} norm appears in all methods.
This phenomenon is revisited in Section~\ref{sec:secondorder}

The \ac{wussadl} heuristic appears twice in these tables: in the first case, its whitening transformation is obtained using statistical linearization (\ac{wussadl}-S), and in the second deterministic linearization (\ac{wussadl}-D).
The unscented transform used for the statistical linearization relies on several parameters to sample points; in this case, the relevant parameter to be adjusted is $\alpha$, which dictates the spread of the selected sigma points.
Here, a relatively large value for $\alpha$ of $0.5$ is chosen to spread the sigma points and amplify potential differences between the two linearizations caused by nonlocal nonlinearities.
Despite this, however, the two perform nearly identically in all  cases.
Furthermore, as the metrics compare the performance of the \ac{gm} approximation to a stochastically sampled truth distribution, very small differences between splitting methods are not considered significant with respect to the inherent uncertainties in the metrics incurred from using a sampling-based truth distribution.
\begin{table}[htbp]
    \caption{Molniya case.}
    \label{tab:molniya_baseline_results}
    \centering
    \begin{tabular}{lrrrr}
      \toprule[2pt]%
      Method & MaDEM & CvM norm & MCR\\
      \midrule
      \begin{filecontents*}{data/twobody_molniya_results.csv}
Method,ELK,MHEM,CvMnorm,MCR,Mixands
maxvar,27010258600,66.1355,2968.450,207.8798,27
FOS,1.04E+11,0.1426,145.570,1.6605,27
SOS,1.03E+11,0.1426,145.570,1.6605,27
SOLC,1.03E+11,0.1426,145.570,1.6605,27
SADL,1.41E+11,0.1422,145.120,1.6598,27
US-FOS,67511438531,0.1382,139.540,1.6552,27
US-SOLC,66603156114,0.1382,139.540,1.6552,27
SA-FOS,7476468608,62.0635,2215.320,193.9493,27
SA-SOS,19413238767,54.0678,1860.210,176.7748,27
W-US-SOS,67063296207,0.1382,139.540,1.6552,27
W-US-SOLC,67712163699,0.1382,139.540,1.6552,27
ALoDT,1.04E+11,0.1426,145.570,1.6605,27
W-SA-SOS,15750465358,54.3971,1867.490,177.6046,27
W-US-SADL-S,89262048382,0.1423,145.290,1.6600,27
W-US-SADL-D,89262048382,0.1423,145.290,1.6600,27
\end{filecontents*}
\csvreader[head to column names]{data/twobody_molniya_results.csv}{}
      {\Method & \MHEM & \CvMnorm & \MCR\\}\\[-1em]%
      \bottomrule
    \end{tabular}
\end{table}

The results for the southern $\mathrm{L}_{1}$ halo case are tabulated in Table~\ref{tab:cislunar_baseline_results}.
In this case, uncertainty scaling does provide some significant improvements, with the exception of \ac{sadl}, which seems to perform the best of the base heuristics---those without added uncertainty scaling, spherical-averaging, or whitening.
However, the computational simplicity of the first-order stretching methods ensures that \ac{fos} and \ac{usfos} remain attractive choices for such problems.
Compared to their performances in the previous two cases, the heuristics incorporating spherical-averaging also produce relatively strong results.
Furthermore, unlike with base \ac{fos} and \ac{sos}, extending the spherical-averaging methods to second-order stretching produces notable improvements.
As with the previous cases though, the addition of whitening transforms again makes little impact.
Finally, the entry labeled \textit{original} in the last row of the table reports the difference between the chosen initial Gaussian \ac{pdf} and the particles sampled from it \textit{prior} to any propagation or splitting; thus, any difference between these distributions is solely due to the finite number of samples.
So, distributions whose metrics differ on similar or smaller scales to the sampling impacts are difficult to definitively distinguish.

Figure~\ref{fig:full_page_marginals} offers a qualitative look at the performance of select splitting heuristics, where the depicted marginals can build intuition for the tabulated metrics.
As might be expected from the relatively strong results that most heuristics demonstrated in Table~\ref{tab:GEO_baseline_results} and the relatively linear $x$-$y$ marginal, all three distributions in the \ac{geo} case appear to match the truth quite closely.
The \ac{leo} case largely follows suit; however, here the limitations of linear covariance propagation become more apparent. These limitations manifest visually in the marginal distributions appearing very thin and stretched compared to the truth, and ultimately produced poor structural agreement in the Molniya case.
In the more challenging Molniya and southern $\mathrm{L}_{1}$ halo cases, the \ac{fos}-based splitting fails to appropriately capture the nonlinearities.
Furthermore, as discussed in Section~\ref{sec:secondorder}, extension to second-order propagation of moments produces more accurate \ac{gm} approximations; neglecting these higher-order effects can be severly damaging in highly nonlinear systems.
The marginals provided here are representative, with many of those produced by other splitting heuristics appearing similar.
In particular, the deferred splitting DS-2 and DS-3 are identical to the immediate splitting \ac{wussolc} in the southern $\mathrm{L}_{1}$ halo case, as will be reported in the subsequent subsection.

\begin{table}[htbp]
    \caption{Southern $\mathrm{L}_{1}$ halo case.}
    \label{tab:cislunar_baseline_results}
    \centering
    \begin{tabular}{lrrrr}
      \toprule[2pt]%
      Method & MaDEM & CvM norm & MCR \\
      \midrule
      \begin{filecontents*}{data/twobody_cislunar_754_results.csv}
Method,ELK,MHEM,CvMnorm,MCR,Mixands
maxvar,15153193.8,1.2508,19.9495,9.9705,27
FOS,14264162.67,19.8236,36.0496,246.1854,27
SOS,14278042.43,19.8164,36.0496,246.2240,27
SOLC,14276895.15,19.8199,36.0496,246.2881,27
SADL,28900729.09,0.1461,3.9392,3.2341,27
US-FOS,29302342.89,0.1453,3.7545,3.2109,27
US-SOLC,29320523.57,0.1458,3.7562,3.2106,27
SA-FOS,20605280.72,0.3124,8.4320,4.3848,27
SA-SOS,29277503.46,0.1455,3.7636,3.2132,27
W-US-SOS,29318192.62,0.1457,3.7560,3.2107,27
W-US-SOLC,29316404.54,0.1457,3.7560,3.2106,27
ALoDT,29030086,0.1592,4.1168,3.2375,27
W-SA-SOS,29278463.02,0.1455,3.7634,3.2132,27
W-US-SADL-S,28895437.32,0.1461,3.9392,3.2341,27
W-US-SADL-S,28895437.32,0.1461,3.9392,3.2341,27
original,2115899065,0.0205,0.4376,1.0180,1
\end{filecontents*}
\csvreader[head to column names]{data/twobody_cislunar_754_results.csv}{}
      {\Method & \MHEM & \CvMnorm & \MCR\\}\\[-1em]%
      \bottomrule
    \end{tabular}
\end{table}

\newcommand{\figwidthfactor}{0.273}
\newcommand{\hf}{0.273}
\newcommand{\labelxcoordt}{3.75}
\newcommand{\labelxcoordtt}{3.65}
\newcommand{\example}{cart2polar}
\newcommand{\plotsuffix}{_0_1}
\newcommand{\labelycoord}{-0.7}
\newcommand{\labelycoordt}{-0.5}
\newcommand{\hft}{0.295}
\newcommand{\figwidthfactort}{0.275}
\newcommand{\lw}{0.3}

\begin{figure*}
\centering
\begin{subcaptiongroup}
  \subcaptionlistentry{throwaway}
  \label{fig:\example:truthgeo}
  \begin{tikzpicture}
    \node[anchor=north west] (img) at (0,0)
    {\includesvg[width=\lw\linewidth]{Figures/updated_marginals/geotruth_hist_0_1.svg}};
    \node[text=white] at (4.6,-1.1) {\captiontext*{}};
  \end{tikzpicture}%
  \subcaptionlistentry{throwaway}
  \label{fig:\example:fosgeo}
  \begin{tikzpicture}
    \node[anchor=north west] (img) at (0,0)
    {\includesvg[width=\lw\linewidth]{Figures/updated_marginals/geoFOS_0_1.svg}};
    \node[text=white] at (3.9,-1.1) {\captiontext*{}};
  \end{tikzpicture}
  \subcaptionlistentry{throwaway}
  \label{fig:\example:wusgeo}
  \begin{tikzpicture}
    \node[anchor=north west] (img) at (0,0)
    {\includesvg[width=\lw\linewidth]{Figures/updated_marginals/geoWUSSOLC_0_1.svg}};
    \node[text=white] at (3.9,-1.1) {\captiontext*{}};
  \end{tikzpicture}
  \subcaptionlistentry{throwaway}
  \label{fig:\example:truthleo}
  \begin{tikzpicture}
    \node[anchor=north west] (img) at (0,0)
    {\includesvg[width=\lw\linewidth]{Figures/updated_marginals/leotruth_hist_0_1.svg}};
    \node[text=white] at (4.6,-1.1) {\captiontext*{}};
  \end{tikzpicture}
  \subcaptionlistentry{throwaway}
  \label{fig:\example:fosleo}
  \begin{tikzpicture}
    \node[anchor=north west] (img) at (0,0)
    {\includesvg[width=\lw\linewidth]{Figures/updated_marginals/leoFOS_0_1.svg}};
    \node[text=white] at (3.9,-1.1) {\captiontext*{}};
  \end{tikzpicture}
  \subcaptionlistentry{throwaway}
  \label{fig:\example:wusleo}
  \begin{tikzpicture}
    \node[anchor=north west] (img) at (0,0)
    {\includesvg[width=\lw\linewidth]{Figures/updated_marginals/leoWUSSOLC_0_1.svg}};
    \node[text=white] at (3.9,-1.1) {\captiontext*{}};
  \end{tikzpicture}
  \subcaptionlistentry{throwaway}
  \label{fig:\example:truthmoln}
  \begin{tikzpicture}
    \node[anchor=north west] (img) at (0,0)
    {\includesvg[width=\lw\linewidth]{Figures/updated_marginals/molniyatruth_hist_0_1.svg}};
    \node[text=white] at (4.6,-1.1) {\captiontext*{}};
  \end{tikzpicture}
  \subcaptionlistentry{throwaway}
  \label{fig:\example:fosmoln}
  \begin{tikzpicture}
    \node[anchor=north west] (img) at (0,0)
    {\includesvg[width=\lw\linewidth]{Figures/updated_marginals/molniyaFOS_0_1.svg}};
    \node[text=white] at (3.9,-1.1) {\captiontext*{}};
  \end{tikzpicture}
  \subcaptionlistentry{throwaway}
  \label{fig:\example:wusmoln}
  \begin{tikzpicture}
    \node[anchor=north west] (img) at (0,0)
    {\includesvg[width=\lw\linewidth]{Figures/updated_marginals/molniyaWUSSOLC_0_1.svg}};
    \node[text=white] at (3.9,-1.1) {\captiontext*{}};
  \end{tikzpicture}
  \subcaptionlistentry{throwaway}
  \label{fig:\example:truthcis}
  \begin{tikzpicture}
    \node[anchor=north west] (img) at (0,0)
    {\includesvg[width=\lw\linewidth]{Figures/updated_marginals/cislunar_754truth_hist_0_2.svg}};
    \node[text=white] at (4.6,-1.1) {\captiontext*{}};
  \end{tikzpicture}
  \subcaptionlistentry{throwaway}
  \label{fig:\example:foscis}
  \begin{tikzpicture}
    \node[anchor=north west] (img) at (0,0)
    {\includesvg[width=\lw\linewidth]{Figures/updated_marginals/cislunar_754FOS_0_2.svg}};
    \node[text=white] at (3.9,-1.1) {\captiontext*{}};
  \end{tikzpicture}
  \subcaptionlistentry{throwaway}
  \label{fig:\example:wuscis}
  \begin{tikzpicture}
    \node[anchor=north west] (img) at (0,0)
    {\includesvg[width=\lw\linewidth]{Figures/updated_marginals/cislunar_754WUSSOLC_0_2.svg}};
    \node[text=white] at (3.9,-1.1) {\captiontext*{}};
  \end{tikzpicture}
\end{subcaptiongroup}
\captionsetup{subrefformat=parens}
\caption{Post-propagation marginal distributions in $x-y$ for GEO, LEO, and Molniya cases, $x-z$ for cislunar. Rows divided by case, columns by splitting heuristic; first-order propagation of moments for GEO and LEO, second-order for Molniya and cislunar. \subref{fig:\example:truthgeo} truth for GEO case, \subref{fig:\example:fosgeo} GEO \ac{fos}, \subref{fig:\example:wusgeo} GEO \ac{wussolc}; \subref{fig:\example:truthleo} truth for LEO case, \subref{fig:\example:fosleo} LEO \ac{fos}, \subref{fig:\example:wusleo} LEO \ac{wussolc}; \subref{fig:\example:truthmoln} truth for Molniya case, \subref{fig:\example:fosmoln} Molniya \ac{fos}, \subref{fig:\example:wusmoln} Molniya \ac{wussolc}; \subref{fig:\example:truthcis} truth for cislunar case; \subref{fig:\example:foscis} cislunar \ac{fos}, \subref{fig:\example:wuscis} cislunar \ac{wussolc} (identical to DS-2/3).}
\label{fig:full_page_marginals}
\end{figure*}

\subsection{Deferred Splitting Performance}
The tests described in this subsection focus on deferred splitting performance in the cislunar southern $\mathrm{L}_{1}$ halo case.
Here, the \ac{wussolc} heuristic is used for splitting.
More precisely, the unsquared Frobenius norm appearing in the \ac{wussolc} objective is used as the unweighted splitting criteria, such that the tolerance can be interpreted as a Mahalanobis distance.
At the limits, $\epsilon=0$ results in immediate, non-deferred splitting, whereas $\epsilon\rightarrow \infty$ leads to no splitting at all.
Too small a tolerance results in all splitting being performed immediately---or nearly so---and losing the computational edge over immediate splitting, while too great a tolerance degrades the accuracy of the resulting distribution.
The deferred splitting tests in this subsection are all performed using $\epsilon=0.25$.
The three \ac{hotdogs} variants discussed in Section~\ref{sec:methodology} are employed for deferred splitting here: DS-1, with linear recalculation of \ac{stm}s assuming constant \ac{stt}s; DS-2, with numerical integration for \ac{stm}s at each split but constant \ac{stt}s; and DS-3, with numerical integration for both \ac{stm}s and \ac{stt}s.
The marginals for DS-2 and DS-3 are visually identical to those of \ac{wussolc}, which is presented along with the truth distribution and \ac{fos}-based splitting in Figure~\ref{fig:full_page_marginals}.
There, DS-2 and DS-3 can be seen to qualitatively capture the significant non-Gaussianity well.

The chosen tolerance spreads the splitting times over several days, as a mixand is only split once its \ac{wussolc} reaches the tolerance for the remaining time span.
So, this approach escapes the burden of propagating each mixand over the first days compared to splitting immediately.
At the same time, Table~\ref{tab:cislunar_deferred_results} demonstrates that the end \ac{gm} still approximates the true distribution well, especially with the higher fidelity methods.
Though two of the metrics suggest improved performance over immediate splitting, such small differences likely arise instead from stochastic sampling effects.
The differences between DS-2 or DS-3 and the immediate split are all smaller than those incurred by sampling in Table~\ref{tab:cislunar_baseline_results}, which makes interpretation of these differences difficult.
There is also a noticeable improvement between DS-1 and DS-2, while DS-2 and DS-3 are nearly identical to each other and to the immediate splitting.
This demonstrates that while DS-1 can be safely employed in less stressing scenarios, DS-2 can closely match the immediate splitting for a given number of mixands without necessitating full recomputation of both the \acp{stm} and \acp{stt} at each splitting time: a significant computational saving.

\begin{table}[htbp]
    \caption{Southern $\mathrm{L}_{1}$ halo case, immediate and deferred splitting.}
    \label{tab:cislunar_deferred_results}
    \centering
    \begin{tabular}{lrrrr}
      \toprule[2pt]%
      Method & MaDEM & CvM norm & MCR & Runtime [nd] \\
      \midrule
      \begin{filecontents*}{data/twobody_cislunar_754_deferred_results.csv}
Method,ELK,MHEM,CvMnorm,MCR,Mixands,Runtime,timeit time,,,,,
immediate split,29316404.54,0.1457,3.7560,3.2106,27,1,503.0317918,,,,1,545.7087918
DS-1,3.31E-12,0.2040,3.1390,4.2254,27,0.0605,30.4103926,,,,0.0573,31.246507
DS-2,8.78E-12,0.1528,3.7045,3.2490,27,0.5407,271.9942387,0.5407,,,\textless0.5739,313.1871719
DS-3,8.79E-12,0.1528,3.7044,3.2490,27,0.5590,281.1771487,,,,0.5759,314.27901
\end{filecontents*}
\csvreader[head to column names]{data/twobody_cislunar_754_deferred_results.csv}{}
      {\Method & \MHEM & \CvMnorm & \MCR &\Runtime\\}\\[-1em]%
      \bottomrule
    \end{tabular}
\end{table}

The final column in Table~\ref{tab:cislunar_deferred_results} is the relative runtime of each method nondimensionalized by the runtime for immediate splitting.
Even with a relatively low tolerance, the computational savings of deferred splitting are immediately apparent.
While the savings for DS-1 do come at a slight price in terms of performance, comparison with the other immediate splitting heuristics demonstrates that it still remains competitive with the best results in Table~\ref{tab:cislunar_baseline_results} at a greatly reduced computational cost.
At the same time, DS-2 and DS-3 have negligible differences in their metrics with respect to immediate splitting using \ac{wussolc}, and yet still achieve significantly quicker execution time.
Here, DS-2 and DS-3 also have similar runtimes, despite the approximation of the \acp{stt} as constant.
While employing a \ac{da}-based approach with the DACEyPy library, the difference in computational effort between solely propagating the \acp{stm} and propagating both the \acp{stm} and \acp{stt} is relatively minor, as overhead costs dominate.
In \cite{siciliano2025deferredSplitting}, a variational equation-based approach demonstrated much more dramatic savings between DS-2 and DS-3.
These findings align with other works; in \cite{boone2025EvaluationHigherOrder}, this variational approach exhibited strong performance for lower orders---especially for more complex dynamical systems---while the \ac{da} method showed better scaling as order increased.
In general, however, compared to the runtimes reported for an unperturbed cislunar orbit in \cite{siciliano2025deferredSplitting}, the results presented here suggest that the relative savings are not diminished by the inclusion of higher-fidelity perturbations.
Rather, the fashion in which the partial derivative tensors are computed can determine the degree to which the approximations in the deferred splitting variations affect runtime.

\subsection{Second-Order Moment Propagation Performance}
\label{sec:secondorder}
In Table~\ref{tab:molniya_baseline_results}, the splitting heuristics in the Molniya case exhibited a high \ac{cvm} norm despite relatively strong \ac{madem} and \ac{mcr} results.
This indicated a failure to fully describe the higher-order moments of the true distribution.
In contrast, second-order mixand moment propagation drastically reduces the \ac{cvm} norm, as can be seen in Table~\ref{tab:molniya_second_order_initial}.
Furthermore, the plotted marginal in Figure~\ref{fig:full_page_marginals} qualitatively can be seen to be more diffuse than those of the \ac{geo} and \ac{leo} cases employing first-order propagation of moments.
This observation is supported by other findings that, despite the non-preservation of highest density region volume by higher-order covariance propagation methods, the resulting approximations better fit the true non-Gaussian distributions compared to volume-preserving linear approximations \cite{kulik2025QuantifyingVolumeChanges}.
The \ac{madem} and \ac{mcr} also decrease for each method compared to Table~\ref{tab:molniya_baseline_results}, preserving the general trends between the heuristics.
\begin{table}[htbp]
    \caption{Molniya case, second-order moment propagation.}
    \label{tab:molniya_second_order_initial}
    \centering
    \begin{tabular}{lrrrr}
      \toprule[2pt]%
      Method & MaDEM & CvM norm & MCR \\
      \midrule
      \begin{filecontents*}{data/molniya_results_2ndorder.csv}
Method,ELK,MHEM,CvMnorm,MCR,Mixands
maxvar,1393684.044,57.9700,156.430,176.3610,27
FOS,20130728.86,0.0897,25.926,1.5753,27
SOS,20130864.46,0.0897,25.926,1.5753,27
SOLC,20130863.77,0.0897,25.926,1.5753,27
SADL,20211614.85,0.0895,25.855,1.5749,27
US-FOS,21216343.5,0.0879,24.981,1.5743,27
US-SOLC,21216889.68,0.0879,24.981,1.5743,27
SA-FOS,1434785.205,15.0987,152.583,47.4662,27
SA-SOS,1513485.334,5.9698,145.099,19.8527,27
W-US-SOS,21216172.31,0.0879,24.982,1.5743,27
W-US-SOLC,21216172.33,0.0879,24.982,1.5743,27
ALoDT,20130293.23,0.0897,25.926,1.5753,27
W-SA-SOS,1509942.687,6.0858,145.338,20.2031,27
W-US-SADL-S,20180735.32,0.0896,25.881,1.5750,27
W-US-SADL-D,20180735.32,0.0896,25.881,1.5750,27
\end{filecontents*}
\csvreader[head to column names]{data/molniya_results_2ndorder.csv}{}
      {\Method & \MHEM & \CvMnorm & \MCR\\}\\[-1em]%
      \bottomrule
    \end{tabular}
\end{table}

This incorporation of higher-order terms in the mean and covariance propagation can also be seen in Table~\ref{tab:cislunar_second_order_initial} to lead to significant improvements across all metrics for each splitting method.
With the exception of \ac{safos} and the base \ac{fos}, \ac{sos}, and \ac{solc}, every other heuristic here clearly outperforms even the strongest method in the first-order case from Table~\ref{tab:cislunar_baseline_results}.
Furthermore, the requisite \acp{stt} are already computed for the splitting methods which incorporate second-order effects.
Hence, extension of the propagation to the second order is much more computationally efficient than increasing the number of mixands in the \ac{gm}, as well as more effective.
\begin{table}[htbp]
    \caption{Southern $\mathrm{L}_{1}$ halo case, second-order moment propagation for immediate splitting.}
    \label{tab:cislunar_second_order_initial}
    \centering
    \begin{tabular}{lrrrr}
      \toprule[2pt]%
      Method & MaDEM & CvM norm & MCR \\
      \midrule
      \begin{filecontents*}{data/second_order_twobody_cislunar_754_results.csv}
Method,ELK,MHEM,CvMnorm,MCR,Mixands
maxvar,35358.56024,0.7011,6.6272,7.3088,27
FOS,10398.51945,4.7028,24.9761,37.6789,27
SOS,10397.54262,4.6987,24.9761,37.6557,27
SOLC,10397.63457,4.6988,24.9761,37.6563,27
SADL,232165.6752,0.0981,1.2056,3.0465,27
US-FOS,247834.8274,0.0988,1.1629,3.0378,27
US-SOLC,247792.4354,0.0990,1.1632,3.0378,27
SA-FOS,93797.28246,0.1942,2.4912,3.9917,27
SA-SOS,246816.0295,0.0988,1.1649,3.0386,27
W-US-SOS,247799.9949,0.0990,1.1632,3.0378,27
W-US-SOLC,247799.9619,0.0990,1.1632,3.0378,27
ALoDT,222410.7439,0.1049,1.2469,3.0492,27
W-SA-SOS,246815.2411,0.0988,1.1648,3.0387,27
W-US-SADL-S,232167.2031,0.0981,1.2056,3.0465,27
W-US-SADL-D,232167.2031,0.0981,1.2056,3.0465,27
\end{filecontents*}
\csvreader[head to column names]{data/second_order_twobody_cislunar_754_results.csv}{}
      {\Method & \MHEM & \CvMnorm & \MCR\\}\\[-1em]%
      \bottomrule
    \end{tabular}
\end{table}

The second-order moment propagation is also applied to the deferred splitting case, with results presented in Table~\ref{tab:cislunar_second_order_deferred}.
As with several of the immediate splitting methods, the \ac{hotdogs} algorithm eagerly computes the \acp{stt}.
Thus, the incorporation of the higher-accuracy second-order mixand moment propagation in \ac{hotdogs} still retains its notable computational savings over immediate splitting, especially in DS-1 and DS-2.
DS-3 shows only a modest reduction in runtime over the immediate splitting reference; however, there remains room for optimization in the supplemental computations---e.g., finding the splitting time.
Thus, this value should be interpreted more as an upper bound.
While the \ac{cvm} norm drops sharply across all listed methods, the \ac{madem} and \ac{mcr} performance gains for DS-1 and DS-2 are relatively small; however, DS-3 sees significant improvement in both over the linear propagation case.
Furthermore, when in the linear case DS-2 and DS-3 both matched the immediate splitting performance well, DS-3 showed little benefit over DS-2---especially when considering the computational cost.
Here, however, DS-3 noticeably differentiates itself from DS-2 and maintains close performance to the immediate split.
This indicates that the full \ac{stt} recomputation found in DS-3 allows it to better capture higher-order effects emerging from the higher-order propagation than DS-1 or DS-2.
\begin{table}[htbp]
    \caption{Southern $\mathrm{L}_{1}$ halo case, second-order moment propagation for deferred splitting.}
    \label{tab:cislunar_second_order_deferred}
    \centering
    \begin{tabular}{lrrrr}
      \toprule[2pt]%
      Method & MaDEM & CvM norm & MCR & Runtime [nd] \\
      \midrule
      \begin{filecontents*}{data/second_order_twobody_cislunar_754_deferred_results.csv}
Method,ELK,MHEM,CvMnorm,MCR,Mixands,Runtime,timeit time,,,,,,,,
immediate split,247799.9619,0.0990,1.1632,3.0378,27.0000,1,491.6826941,,,,513.0740972,,,,
DS-1,2.74E-11,0.2554,2.5326,4.2057,27.0000,0.0868,42.6627251,,,,42.0942951,,,,
DS-2,4.00E-11,0.1899,1.5929,3.2389,27.0000,0.8178,402.1201827,,,,466.5427964,,0.0918,47.1019866,x10
DS-3,6.78E-11,0.1033,1.1639,3.0729,27.0000,0.9205,452.6001462,,0.8178,,467.7867787,,,,
\end{filecontents*}
\csvreader[head to column names]{data/second_order_twobody_cislunar_754_deferred_results.csv}{}
      {\Method & \MHEM & \CvMnorm & \MCR &\Runtime\\}\\[-1em]%
      \bottomrule
    \end{tabular}
\end{table}

\subsection{Nonlinearity Analysis}
\begin{figure}[htbp]
\centering
{
  \ifimporttikz
    \tikzsetnextfilename{wussolcs}
    \import{Figures/}{wussolcs.tikz}
  \else
    \includegraphics{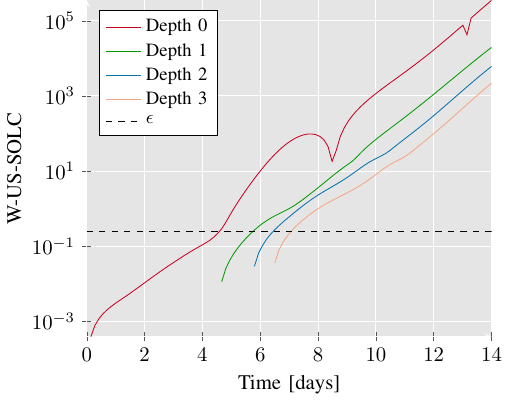}
  \fi
}
\caption{Weighted \ac{wussolc} criterion for DS-3 variant in southern $\mathrm{L}_{1}$ halo orbit case.}
\label{fig:cislunar_deferred_3_tol}
\end{figure}

Using second-order propagation of the mean and covariance, Figure~\ref{fig:cislunar_deferred_3_tol} plots the \ac{wussolc} unsquared Frobenius norm scaled by mixand weight over 14 days for the original mixand at depth zero, as well as for the central child mixands at split depths of one, two, and three.
The performance benefits of deferred splitting from reducing the time span over which the child mixands must be propagated are depicted in the delayed emergence of the higher depth curves, which appear after the preceding depth's \ac{wussolc} meets the tolerance of $0.25$.
It should be noted that \ac{wussolc}, as a measure of the system's nonlinearities, is not strictly increasing in time, which informs the selection of splitting time as the last occasion on which the objective functional hits the splitting tolerance.
This prevents potential wasted effort propagating mixands if splitting occurs before a reduction in \ac{wussolc} back to acceptable values.

The motivation for Gaussian splitting is to reduce the covariances of the mixands, and thereby decrease the impact of nonlinearities.
As such, a measure of the nonlinearity impact imparted on a post-split mixand should in principle be strictly less than that of the original component over the propagation period.
Whereas recomputing the whitening transformation after each split may counterintuitively result in larger \ac{wussolc} despite improving the uncertainty propagation accuracy \cite{kulik2024NonlinearityUncertaintyInformed}, the strategy proposed in Section~\ref{sec:methodology} of this work of employing a fixed reference whitening transformation is seen in Figure~\ref{fig:cislunar_deferred_3_tol} to result in monotonically decreasing \ac{wussolc} with increasing split depth.

\section{Conclusion}
\label{sec:conclusion}
\acresetall
This paper introduces \ac{hotdogs}, a novel algorithm for deferred Gaussian splitting utilizing the \acp{stm} and \acp{stt} of the flow-map.
This enables significant computational savings by shortening the propagation time for most mixands while retaining an accurate representation of the true probability distribution.
Results for uncertainty propagation using Gaussian mixtures split at the starting time are first presented for four cases representative of typical two- and multi-body dynamical problems: a geostationary orbit, a low Earth orbit, a Molniya orbit, and an Earth-Moon southern $\mathrm{L}_{1}$ halo orbit.
Analysis of the applied splitting heuristics emphasizes the importance of uncertainty scaling in selection of splitting direction in multi-body systems as well as the suitability of \ac{fos}-based methods in astrodynamical contexts.
Then, three variants of \ac{hotdogs} offering different fidelity improvements and computational complexity are applied to uncertainty propagation in the challenging three-body orbit case.
The deferred splitting algorithm demonstrates performance indistinguishable from the best immediate splitting methods with a reduced computational burden for two of the three variants.
The other variant boasted a $17\times$ faster runtime than the baseline case while only incurring slightly higher errors than the other variants.
Extension to second-order moment propagation saw further improvements in all three measures of accuracy considered.
Additionally, utilizing the eagerly computed \acp{stm} and \acp{stt} enabled up to $12\times$ savings in runtime over the immediate splitting approach.
This establishes the potential for advantageous implementation of the fastest method in less stressing or computationally constrained scenarios, while the other two exhibit suitability for benefiting an even broader range of applications.

\section*{Acknowledgment}
 The authors would like to thank Dr.\ Andrea De Vittori, who implemented the high-fidelity force models used in the analysis.

\appendix
\subsection{Shift Calculation for Symmetric Higher-Order Power Iteration}
\label{sec:shift_calculation}
In order to guarantee convergence of symmetric shifted higher-order power iteration for the tensor eigenvalue calculation upon which \ac{sos} and \ac{wussos} depend \cite{kulik2024NonlinearityUncertaintyInformed}, an appropriate shift factor must be computed. According to \cite{kolda2011ShiftedPowerMethod}, if $\mathbf{T}$ is an order $m$ symmetric tensor, then the shift factor $\eta$ must satisfy
\begin{equation}
    \eta\geq (m-1)\max_{\Vert\mathbf{x}\Vert_2=1} \rho(\mathbf{T}\mathbf{x}^{m-2})
\end{equation}
in order to guarantee convergence of shifted higher-order power iteration, where $\rho$ denotes the spectral radius of the matrix given by contracting $\mathbf{T}$ with $m-2$ copies of $\mathbf{x}$. One conservative choice of the shift parameter is given by \cite{kolda2011ShiftedPowerMethod}
\begin{equation}
\eta^*=(m-1)\sum_{i_1,...,i_m} \vert \mathbf{T}_{i_1,...,i_m}\vert
\end{equation}
However, when this shift parameter is used for calculation of \ac{wussos} and \ac{sos}, the convergence can be prohibitively slow. Thus, a  less conservative choice of shift parameter that still guarantees convergence is derived here.
First, note that
\begin{align}
\max_{\Vert\mathbf{x}\Vert_2=1} \rho(\mathbf{T}\mathbf{x}^{m-2}) \leq \max_{\Vert\mathbf{x}\Vert_2=1} \Vert\mathbf{T}\mathbf{x}^{m-2}\Vert_F
\end{align}
because $\mathbf{T}\mathbf{x}^{m-2}$ is a symmetric matrix and the spectral radius is equal to the induced 2-norm for a symmetric matrix which is bounded above by the Frobenius norm.
Further,
\begin{align}
\max_{\Vert\mathbf{x}\Vert_2=1} \Vert\mathbf{T}\mathbf{x}^{m-2}\Vert_F = \max_{\Vert\mathbf{x}\Vert_2=1} \Vert\mathbf{T}^{'}\mathbf{x}^{m-2}\Vert_2\leq\max_{\Vert\mathbf{y}\Vert_2=1} \Vert\mathbf{T}^{''}\mathbf{y}\Vert_2
\end{align}
where $\mathbf{T}^{'}$ is an  $n^2\times \underbrace{n \times ...\times n}_{m-2\text{ times}}$ tensor flattening of $\mathbf{T}$ while maintaining the order of indices and $\mathbf{T}^{''}$ is an  $n^2\times n^{m-2}$ matrix where the last $m-2$ axes of $\mathbf{T}^{'}$ are flattened into one axis. The final inequality comes from recognizing that the middle term is the induced 2-norm which is bounded above by the induced 2-norm of the matrix which flattens all of the final axes into one single axis of the matrix.
See the upper bound on the induced 2-norm in \cite{kulik2024applications}.
As such, the maximum singular value of the matrix $\mathbf{T}^{''}$ is employed in the shift parameter:
\begin{equation}
    \eta^{**}=(m-1)\Vert\mathbf{T}^{''}\Vert_2
\end{equation}
Clearly, $\eta^{**}\leq\eta^{*}$ since \begin{equation}
    \Vert\mathbf{T}^{''}\Vert_2\leq\Vert \mathbf{T}^{''}\Vert_F=\Vert\mathrm{vec}(\mathbf{T}^{''})\Vert_2\leq \Vert\mathrm{vec}(\mathbf{T}^{''})\Vert_1
\end{equation}
and thus convergence should be accelerated when employing the shift parameter $\eta^{**}$ at the cost of calculating a matrix 2-norm once prior to the shifted higher-order power iteration algorithm.

\subsection{Role of Whitening Transformation in Measure Behavior} \label{sec:whitening_ex}
Splitting of a Gaussian component results in strictly decreasing covariance.
Therefore, any measure of the impacts of nonlinearity on a mixand---such as the splitting objective functionals presented in this paper---should be monotonically decreasing in split depth; however, those with whitening transformations following \cite{kulik2024NonlinearityUncertaintyInformed} do not always adhere to this.
To illustrate the potential impacts of linearization errors in the whitening transformation, consider a Gaussian with identity covariance and unity \ac{wussos} prior to splitting.
\begin{align}
    \mathbf{P}(t_0) = \begin{bmatrix}
        1 & 0\\ 0 & 1
    \end{bmatrix}\\
    \mathcal{F}_{\text{\ac{wussos}}}(g, \delta) = 1
\end{align}
The subscript ``\ac{wussos}'' is used to distinguish the specific objective functional considered here.
Let the \ac{stm} and \ac{stt} be given by
\begin{align}
    \mathbf{\Phi} &= \begin{bmatrix}
        1 & 0\\ 0 & 1
    \end{bmatrix}\\
    \Psi^1_{j, k} &= \begin{bmatrix}
        0 & 0\\ 0 & 1
    \end{bmatrix}, \quad \Psi^2_{j, k} = \begin{bmatrix}
        1 & 0\\ 0 & 0
    \end{bmatrix}
\end{align}

With an isotropic covariance, an arbitrary direction can be chosen for splitting.
The resulting mixands must have lower variance along the splitting direction; so, selecting the $x_{1}$-axis for splitting, the $i$th child mixand will have covariance
\begin{align}
    \mathbf{P}_i(t_0) &= \mathbf{P}(t_0) - \beta \begin{bmatrix}
        1 \\ 0
    \end{bmatrix} \begin{bmatrix}
        1 & 0
    \end{bmatrix}, \quad 0 < \beta < 1\\
    \mathbf{P}_i(t_0) &= \begin{bmatrix}
        1 - \beta & 0\\ 0 & 1
    \end{bmatrix}
\end{align}
where $\beta$ is some scaling factor that accounts for the reduction of the original covariance in the direction of the split.
With the linear covariance approximation, the propagated post-split covariance becomes
\begin{align}
    \mathbf{P}_i(t_f) &= \mathbf{\Phi} \mathbf{P}_i(t_0) \mathbf{\Phi}^\top\\
    \mathbf{P}_i(t_f) &= \begin{bmatrix}
        1 - \beta & 0\\ 0 & 1
    \end{bmatrix}
\end{align}

After splitting along the $x_{1}$-axis, the next candidate split direction---the direction which maximizes the objective functional---can be shown to intuitively lie along the $x_{2}$-axis: $\delta^* = \begin{bmatrix}
    0&1
\end{bmatrix}^\top$. As a potential splitting direction, $\delta^*$ can be seen to satisfy the constraint on \ac{wussos}:
\begin{align}
    \Vert \delta^{*} \Vert_{\mathbf{P}_x^{-1}} =
    \begin{bmatrix}
        0 & 1
    \end{bmatrix} \begin{bmatrix}
        1 - \beta & 0\\ 0 & 1
    \end{bmatrix}^{-1} \begin{bmatrix}
        0 \\ 1
    \end{bmatrix} &= 1
\end{align}
The functional then becomes
\begin{align}
    \mathcal{F}_{\text{\ac{wussos}}}(\mathbf{g},\mathbf{\bm{\delta}}; \Theta) = \Vert \mathbf{\Psi} \delta^{*^2} \Vert_{\mathbf{P}_i^{-1/2}(t_f)} \\
    = \sqrt{(\mathbf{\Psi} \delta^{*^2})^\top \mathbf{P}_i^{-1/2}(t_f) \mathbf{\Psi} \delta^{*^2} }
\end{align}
Substituting in the values for the \ac{stt} and $\delta^{*}$,
\begin{align}
    \mathbf{\Psi} \delta^{*^2} = \begin{bmatrix}
        1 \\ 0
    \end{bmatrix}
\end{align}
So, the \ac{wussos} for the post-split mixand evaluates to
\begin{align}
    \mathcal{F}_{\text{\ac{wussos}}}(\mathbf{g},\mathbf{\bm{\delta}}; \Theta) &= \sqrt{\begin{bmatrix}
        1 & 0
    \end{bmatrix} \begin{bmatrix}
        1-\beta & 0\\ 0 & 1
    \end{bmatrix} \begin{bmatrix}
        1 \\ 0
    \end{bmatrix}}\\
    \mathcal{F}_{\text{\ac{wussos}}}(\mathbf{g},\mathbf{\bm{\delta}}; \Theta) &= \sqrt{\frac{1}{1 - \beta}} > 1
\end{align}

As $\beta$ is defined to be between zero and one, using the post-split output covariance for whitening would ultimately increase \ac{wussos} following the split in this case.

\balance
\bibliographystyle{unsrt}
\bibliography{kl_zotero_bib}

\begin{thebibliography}{10}

\bibitem{adurthi2015ConjugateUnscentedTransformationBased}
Nagavenkat Adurthi and Puneet Singla.
\newblock Conjugate {{Unscented Transformation-Based Approach}} for {{Accurate Conjunction Analysis}}.
\newblock {\em Journal of Guidance, Control, and Dynamics}, 38(9):1642--1658, September 2015.

\bibitem{stojanovski2021HigherOrderUnscentedEstimator}
Zvonimir Stojanovski and Dmitry Savransky.
\newblock Higher-{{Order Unscented Estimator}}.
\newblock {\em Journal of Guidance, Control, and Dynamics}, 44(12):2186--2198, 2021.

\bibitem{park2006NonlinearMappingGaussian}
Ryan~S. Park and Daniel~J. Scheeres.
\newblock Nonlinear {{Mapping}} of {{Gaussian Statistics}}: {{Theory}} and {{Applications}} to {{Spacecraft Trajectory Design}}.
\newblock {\em Journal of Guidance, Control, and Dynamics}, 29(6):1367--1375, 2006.

\bibitem{majji2008high}
Manoranjan Majji, John~L Junkins, and James~D Turner.
\newblock A high order method for estimation of dynamic systems.
\newblock {\em The Journal of the Astronautical Sciences}, 56(3):401--440, 2008.

\bibitem{boone2024efficient}
Spencer Boone and Jay McMahon.
\newblock An efficient approximation of the second-order extended kalman filter for a class of nonlinear systems.
\newblock In {\em 2024 European Control Conference (ECC)}, pages 3533--3538. IEEE, 2024.

\bibitem{servadio2020recursive}
Simone Servadio and Renato Zanetti.
\newblock Recursive polynomial minimum mean-square error estimation with applications to orbit determination.
\newblock {\em Journal of Guidance, Control, and Dynamics}, 43(5):939--954, 2020.

\bibitem{servadio2022MaximumPosterioriEstimation}
Simone Servadio, Renato Zanetti, and Roberto Armellin.
\newblock Maximum {{A Posteriori Estimation}} of {{Hamiltonian Systems}} with {{High Order Taylor Polynomials}}.
\newblock {\em The Journal of the Astronautical Sciences}, 69(2):511--536, April 2022.

\bibitem{acciarini2024NonlinearPropagationNonGaussian}
Giacomo Acciarini, Nicola Baresi, David J.~B. Lloyd, and Dario Izzo.
\newblock Nonlinear {{Propagation}} of {{Non-Gaussian Uncertainties}}, November 2024.

\bibitem{horwood2014GaussMisesDistribution}
Joshua~T. Horwood and Aubrey~B. Poore.
\newblock Gauss von {{Mises Distribution}} for {{Improved Uncertainty Realism}} in {{Space Situational Awareness}}.
\newblock {\em SIAM/ASA Journal on Uncertainty Quantification}, 2(1):276--304, January 2014.

\bibitem{jones2019MultifidelityOrbitUncertainty}
Brandon~A. Jones and Ryan~M. Weisman.
\newblock Multi-fidelity orbit uncertainty propagation.
\newblock {\em Acta Astronautica}, 155:406--417, February 2019.

\bibitem{jones2013NonlinearPropagationOrbit}
Brandon~A Jones, Alireza Doostan, and George~H Born.
\newblock Nonlinear {{Propagation}} of {{Orbit Uncertainty Using Non-Intrusive Polynomial Chaos}}.
\newblock {\em Journal of Guidance, Control \& Dynamics}, 36(2):430, March 2013.

\bibitem{jones2023incorporating}
Brandon~A Jones and Trevor~N Wolf.
\newblock Incorporating directional uncertainties into polynomial chaos expansions for astronautics problems.
\newblock {\em The Journal of the Astronautical Sciences}, 70(4):19, 2023.

\bibitem{vittaldev2016SpacecraftUncertaintyPropagation}
Vivek Vittaldev, Ryan~P. Russell, and Richard Linares.
\newblock Spacecraft {{Uncertainty Propagation Using Gaussian Mixture Models}} and {{Polynomial Chaos Expansions}}.
\newblock {\em Journal of Guidance, Control, and Dynamics}, 39(12):2615--2626, 2016.

\bibitem{acciarini2024uncertainty}
Giacomo Acciarini, Cristian Greco, and Massimiliano Vasile.
\newblock Uncertainty propagation in orbital dynamics via {G}alerkin projection of the {Fokker-Planck} equation.
\newblock {\em Advances in Space Research}, 73(1):53--63, 2024.

\bibitem{kumar2012nonlinear}
Mrinal Kumar and Suman Chakravorty.
\newblock Nonlinear filter based on the {Fokker-Planck} equation.
\newblock {\em Journal of guidance, control, and dynamics}, 35(1):68--79, 2012.

\bibitem{adurthi2022estimation}
Naga~Venkat Adurthi and Manoranjan Majji.
\newblock Estimation of dynamic systems using a method of characteristics filter.
\newblock {\em Automatica}, 140:110226, 2022.

\bibitem{khatri2024HybridMethodUncertainty}
Yashica Khatri and Daniel~J. Scheeres.
\newblock Hybrid {{Method}} of {{Uncertainty Propagation}} for {{Near-Earth Conjunction Analysis}}.
\newblock {\em Journal of Guidance, Control, and Dynamics}, pages 1--14, June 2024.

\bibitem{huber2008entropy}
Marco~F. Huber, Tim Bailey, Hugh~F. {Durrant-Whyte}, and Uwe~D. Hanebeck.
\newblock On entropy approximation for {{Gaussian}} mixture random vectors.
\newblock {\em IEEE International Conference on Multisensor Fusion and Integration for Intelligent Systems}, pages 181--188, 2008.

\bibitem{demars2013EntropyBasedApproachUncertainty}
Kyle~J. DeMars, Robert~H. Bishop, and Moriba~K. Jah.
\newblock Entropy-{{Based Approach}} for {{Uncertainty Propagation}} of {{Nonlinear Dynamical Systems}}.
\newblock {\em Journal of Guidance, Control, and Dynamics}, 36(4):1047--1057, 2013.

\bibitem{legrand2023BayesianAnglesOnlyCislunar}
Keith~A. LeGrand, Aneesh~V. Khilnani, and John~L. Iannamorelli.
\newblock Bayesian {{Angles-Only Cislunar Space Object Tracking}}.
\newblock In {\em Proceedings of the 33rd {{AAS}}/{{AIAA Space Flight Mechanics Meeting}}}, January 2023.

\bibitem{iannamorelli2025AdaptiveGaussianMixture}
John~L. Iannamorelli and Keith~A. LeGrand.
\newblock Adaptive {{Gaussian Mixture Filtering}} for {{Multi-sensor Maneuvering Cislunar Space Object Tracking}}.
\newblock {\em The Journal of the Astronautical Sciences}, 72(2):2, January 2025.

\bibitem{legrand2022SplitHappensImprecise}
Keith~A. LeGrand and Silvia Ferrari.
\newblock Split {{Happens}}! {{Imprecise}} and {{Negative Information}} in {{Gaussian Mixture Random Finite Set Filtering}}.
\newblock {\em Journal of Advances in Information Fusion}, 17(2):78--96, December 2022.

\bibitem{tuggle2018AutomatedSplittingGaussian}
Kirsten Tuggle and Renato Zanetti.
\newblock Automated {{Splitting Gaussian Mixture Nonlinear Measurement Update}}.
\newblock {\em Journal of Guidance, Control, and Dynamics}, 41(3), 2018.

\bibitem{losacco2024LowOrderAutomaticDomain}
Matteo Losacco, Alberto Foss{\`a}, and Roberto Armellin.
\newblock Low-{{Order Automatic Domain Splitting Approach}} for {{Nonlinear Uncertainty Mapping}}.
\newblock {\em Journal of Guidance, Control, and Dynamics}, 47(2):291--310, 2024.

\bibitem{vittaldevMultidirectionalGaussianMixture}
Vivek Vittaldev and Ryan~P. Russell.
\newblock Multidirectional {{Gaussian Mixture Models}} for {{Nonlinear Uncertainty Propagation}}.
\newblock {\em Computer Modeling in Engineering \& Sciences}, 2016.

\bibitem{kulik2024NonlinearityUncertaintyInformed}
Jackson Kulik and Keith~A. LeGrand.
\newblock Nonlinearity and {{Uncertainty Informed Moment-Matching Gaussian Mixture Splitting}}.
\newblock {\em (In press) IEEE Transactions on Aerospace and Electronic Systems}, 2025.

\bibitem{kulikLINEARCOVARIANCEFIDELITY}
Jackson Kulik, Braden Hastings, and Keith~A. LeGrand.
\newblock {{Linear covariance fidelity checks and measures of non-{G}aussianity}}.
\newblock In {\em Proceedings of the 2025 AAS/AIAA Space Flight Mechanics Meeting}, 2025.

\bibitem{siciliano2025deferredSplitting}
G.~Andrew Siciliano, Keith~A. LeGrand, and Jackson Kulik.
\newblock Deferred higher-order splitting for adaptive {G}aussian mixture orbit uncertainty propagation.
\newblock In {\em 28th {{International Conference}} on {{Information Fusion}} ({{FUSION}})}, pages 1--8, July 2025.

\bibitem{Katok_Hasselblatt_1995}
Anatole~B. Katok and Boris Hasselblatt.
\newblock {\em Introduction to the Modern Theory of Dynamical Systems}.
\newblock Encyclopedia of Mathematics and its Applications. Cambridge University Press, 1995.

\bibitem{sorenson1971RecursiveBayesianEstimation}
Harold~W. Sorenson and Daniel~L. Alspach.
\newblock Recursive {{Bayesian Estimation}} using {{Gaussian Sums}}.
\newblock {\em Automatica}, 7(4):465--479, 1971.

\bibitem{hanebeck2003ProgressiveBayesianEstimation}
Uwe~D. Hanebeck and Olga Feiermann.
\newblock Progressive {{Bayesian}} estimation for nonlinear discrete-time systems:the filter step for scalar measurements and multidimensional states.
\newblock In {\em 42nd {{IEEE International Conference}} on {{Decision}} and {{Control}} ({{IEEE Cat}}. {{No}}.{{03CH37475}})}, volume~5, pages 5366--5371 Vol.5, December 2003.

\bibitem{huber2011AdaptiveGaussianMixture}
Marco~F. Huber.
\newblock Adaptive {{Gaussian}} mixture filter based on statistical linearization.
\newblock In {\em 14th {{International Conference}} on {{Information Fusion}}}, pages 1--8, 2011.

\bibitem{pellegrini2016computation}
Etienne Pellegrini and Ryan~P. Russell.
\newblock On the computation and accuracy of trajectory state transition matrices.
\newblock {\em Journal of Guidance, Control, and Dynamics}, 39(11):2485--2499, 2016.

\bibitem{valli2013nonlinear}
Monica Valli, Roberto Armellin, Pierluigi Di~Lizia, and Michele~R. Lavagna.
\newblock Nonlinear mapping of uncertainties in celestial mechanics.
\newblock {\em Journal of Guidance, Control, and Dynamics}, 36(1):48--63, 2013.

\bibitem{tuggle2020ModelSelectionGaussian}
Kirsten Tuggle.
\newblock {\em Model {{Selection}} for {{Gaussian Mixture Model Filtering}} and {{Sensor Scheduling}}}.
\newblock PhD thesis, University of Texas at Austin, 2020.

\bibitem{doi:10.2514/6.2018-0398}
Mauro Massari, Pierluigi~Di Lizia, Francesco Cavenago, and Alexander Wittig.
\newblock {\em Differential Algebra software library with automatic code generation for space embedded applications}.

\bibitem{Curtis2014Chapter12}
Howard~D. Curtis.
\newblock Chapter 12 - introduction to orbital perturbations.
\newblock In Howard~D. Curtis, editor, {\em Orbital Mechanics for Engineering Students}, pages 651--720. Butterworth-Heinemann, third edition, 2014.

\bibitem{faubel2010FurtherImprovementAdaptive}
Friedrich Faubel and Dietrich Klakow.
\newblock Further improvement of the adaptive level of detail transform: {{Splitting}} in direction of the nonlinearity.
\newblock In {\em 2010 18th {{European Signal Processing Conference}}}, pages 850--854, August 2010.

\bibitem{boone2025EvaluationHigherOrder}
Spencer Boone, Roberto Armellin, Thomas Caleb, Jay McMahon, and St\'ephanie Lizy-Destrez.
\newblock {Evaluation} {and} {Comparison} {of} {Higher}-{Order} {Methods} {in} {Astrodynamics}.
\newblock In {\em Proceedings of the 2025 {{AIAA}}/{{AAS Space Flight Mechanics Meeting}}}, 2025.

\bibitem{kulik2025QuantifyingVolumeChanges}
Jackson Kulik and Keith~A. LeGrand.
\newblock Quantifying volume changes from covariance propagation.
\newblock In {\em 2025 AAS/AIAA Astrodynamics Specialist Conference}, 2025.

\bibitem{kolda2011ShiftedPowerMethod}
Tamara~G. Kolda and Jackson~R. Mayo.
\newblock Shifted {{Power Method}} for {{Computing Tensor Eigenpairs}}.
\newblock {\em SIAM Journal on Matrix Analysis and Applications}, 32(4):1095--1124, October 2011.

\bibitem{kulik2024applications}
Jackson Kulik, Cedric Orton-Urbina, Maximilian~E. Ruth, and Dmitry Savransky.
\newblock Applications of induced tensor norms to guidance navigation and control.
\newblock {\em (In press) Journal of Guidance, Controls, and Dynamics}, 2025.

\end{thebibliography}

\end{document}